\documentclass[reprint,amsmath,amssymb,aps,floatfix]{revtex4-2}

\let\vec\mathbf

\usepackage{graphicx}
\usepackage{dcolumn}
\usepackage{bm}

\usepackage[dvipsnames]{xcolor}
\usepackage{soul}

\usepackage[hidelinks]{hyperref}
\usepackage{amsmath,amssymb}
\usepackage{bbm}
\usepackage{placeins}
\usepackage{tabularx}

\begin{document}

\title{Measuring local moiré lattice heterogeneity of twisted bilayer graphene}

\author{Tjerk Benschop$^{1*}$}
\author{Tobias A. de Jong$^{1*}$}
\author{Petr Stepanov$^{2*}$}
\author{Xiaobo Lu$^{2}$}
\author{Vincent Stalman$^{1}$}
\author{Sense Jan van der Molen$^{1}$}
\author{Dmitri K. Efetov$^{2}$}
\author{Milan P. Allan$^{1}$}
\affiliation{\phantom{b} \\ $^1$ Leiden Institute of Physics, Leiden University, Niels Bohrweg 2, 2333 CA Leiden, the Netherlands}
\affiliation{$^2$ ICFO – Institut de Ciencies Fotoniques, The Barcelona Institute of Science and Technology,\\ Castelldefels, Barcelona, Spain}
\affiliation{\phantom{b} \\ $^*$ : These authors contributed equally to the work.\\
Submitted and accepted in Phys.\ Rev.\ Research.}


\begin{abstract}
\noindent We introduce a new method to continuously map inhomogeneities of a moir\'e lattice and apply it to large-area topographic images we measure on open-device twisted bilayer graphene (TBG). We show that the variation in the twist angle of a TBG device, which is frequently conjectured to be the reason for differences between devices with a supposed similar twist angle, is about 0.08$^\circ$ around the average of 2.02$^\circ$ over areas of several hundred nm, comparable to devices encapsulated between hBN slabs. We distinguish between an effective twist angle and local anisotropy and relate the latter to heterostrain. Our results imply that for our devices, twist angle heterogeneity has a roughly equal effect to the electronic structure as local strain. The method introduced here is applicable to results from different imaging techniques, and on different moir\'e materials. 
\end{abstract}

\maketitle


\section{Introduction}
Stacking two sheets of identical periodic lattices with a small twist angle $\theta$ leads to a super-periodic lattice with moir\'e lattice constant $\lambda(\theta)$ much larger than the original lattice constant a (figure 1a). This new lattice is called a moir\'e lattice. When using atomic layers exfoliated from van der Waals materials, and stacking them with a twist angle, the electronic and structural properties are modulated on the moir\'e length scale $\lambda(\theta)$, leading to the potential for new, emergent electronic properties of the moir\'e material \cite{n1,n2}.\\
Such new properties have been spectacularly demonstrated in twisted bilayer graphene (TBG) around the magic angle of $\theta \approx 1.1 ^\circ$ \cite{n3,n4,n5,n6,n7,n8,n9,n10,n11}. In TBG, the moir\'e lattice modulates the interlayer coupling between the individual graphene sheets, as well as the van der Waals forces on the individual carbon atoms. The former leads to flat bands of low-kinetic-energy electrons \cite{n1}. The latter leads to a slight deformation of the graphene lattice and bandgaps that separate the more localized electrons from the other bands \cite{n1}. When the flat bands are tuned to the Fermi level, they pair and condense into a superfluid at temperatures much higher than what one would naively expect at the low carrier densities observed in TBG \cite{n4}. Additionally, a variety of insulating and metallic behavior has been observed in TBG for different twist angles and band-fillings \cite{n3,n5,n6,n12}.\\
The kinetic energy of the electrons changes rapidly as the twist angle is varied, especially around the magic angle, therefore the fabrication of devices with just the right angle is key in making them superconducting. But getting the right angle might not even be the most challenging aspect of fabricating high-quality superconducting TBG devices: contaminations, internal stress, and heterogeneities of the twist angle are difficult to avoid. This is in part because the magic angle is not the lowest energy configuration and in part because of the strong forces associated with the tear-and-stack technique. Internal stress and heterogeneities are often conjectured to limit the quality of the devices and are attributed as the main causes for the variability between devices \cite{n13}. This holds especially for open devices that lack the hBN top layer; notably such devices have never been found to superconduct.  Measuring, visualizing, and characterizing heterogeneity in the twist angle and strain in TBG is thus crucial to understand and improve devices.\\
Probably the most complete visualization of inhomogeneity thus far has been obtained using scanning SQUID-on-tip microscopy (SOT) \cite{n14}.  SOT measures the Landau levels as a function of location and thus has access to the local superlattice carrier density. On encapsulated devices, SOT has been used to visualize heterogeneity on length scales of a few micron with a resolution of several tens of nanometers, demonstrating that the twist angle varies by less than 4\% \cite{n14}. While being a very precise measure of the local twist angle, SOT is also influenced by other factors, e.g. inhomogeneities of the chemical potential and the local magnetic screening. Other techniques to access homogeneity are Nano-ARPES \cite{n15,n16,n17}, which can image the full electronic structure in reciprocal space with a spatial resolution of circa 600 nm, low energy electron microscopy (LEEM) \cite{n15}, which can image structural inhomogeneities at twist angles lower and higher than the magic angle, conductive atomic-force microscopy (AFM) \cite{n18}, nano-photocurrent mapping \cite{n19}, which can measure the twist angle with a resolution on the order of ~20 nm, and scanning single electron transistors \cite{n20}, which can map the twist angle by measuring the inverse local compressibility. Finally, scanning tunneling microscopy (STM), the probe used in this study, has been used to measure both the topography and the local density of states of TBG, including the emergence of correlations at the magic angle \cite{n21,n22,n23,n24,n25,n26}.

\begin{figure}
\centering
\includegraphics[width=\linewidth]{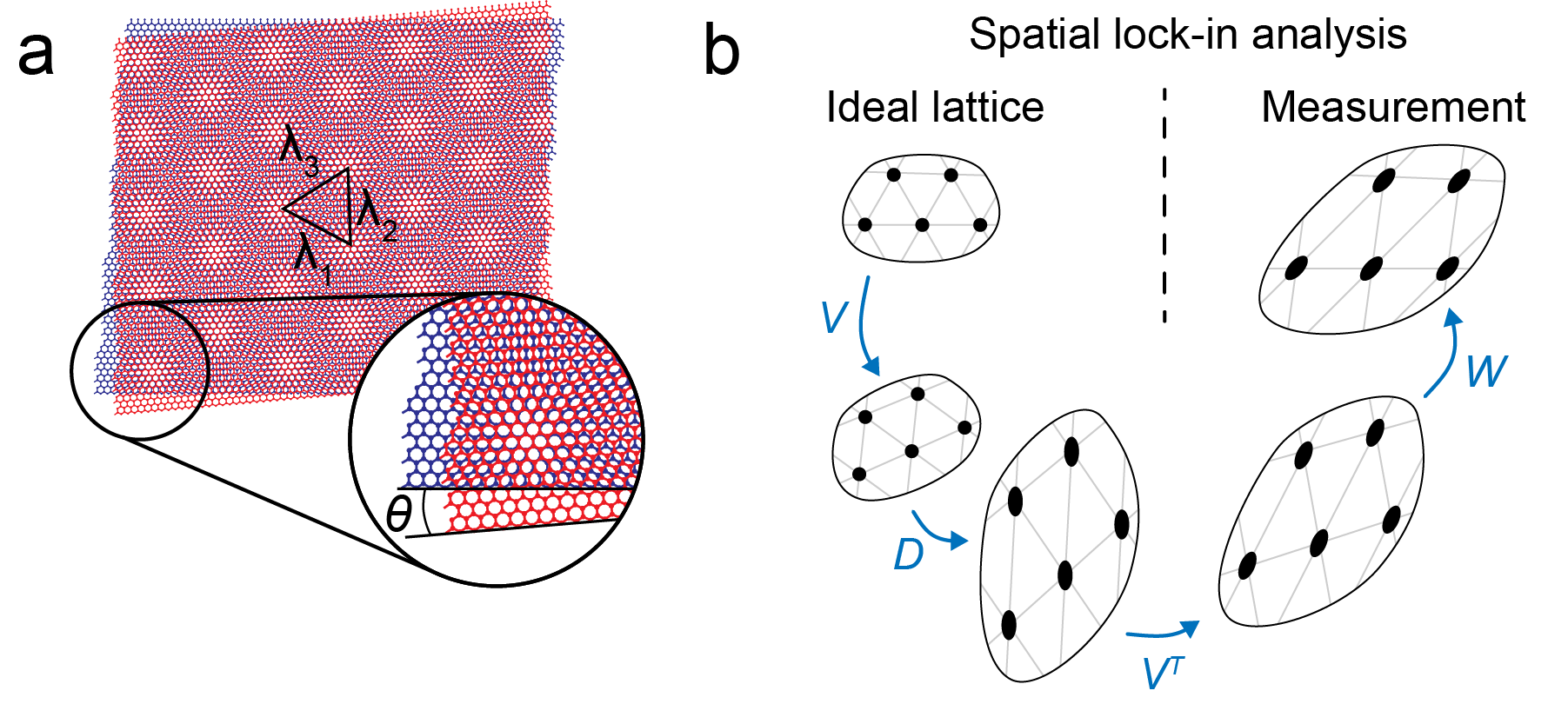}
\caption{a) Moir\'e pattern created by stacking 2 hexagonal lattices with a twist angle of 5$^\circ$. b) Schematic representation of our spatial lock-in algorithm to map the local twist angle and anisotropy.  The measured lattice can be thought of as the result of a series of transformations applied to an ideal lattice. The scaling transformation, $D$, holds information about the local twist angle $\theta^*(\vec{r})$ and the intrinsic local strain present in the device, $\kappa(\vec{r})$. V gives the direction of this local strain, $\psi(\vec{r})$. Finally, $W$ indicates the relative angle between the bilayer and the underlying hBN substrate.}
\end{figure}

In previous STM studies, two different methods have been used to determine the local twist angle. First, one can determine the twist angle using three neighboring moir\'e lattice sites in real space. The distances between each lattice site, $\lambda_1, \lambda_2, \lambda_3$ are fit to a set of equations that yield the twist angle at a per-triangle resolution (figure 1a) \cite{n22}, and, using a model with assumptions about the strain distribution in the two layers, an estimate for the heterostrain $\epsilon$.\\
A second method to determine the twist angle uses the Fourier transform of a real space topography to determine the moir\'e wavelengths $\lambda_j$ in the three directions of the moir\'e lattice (in principle, two directions are fully determining the lattice, but often all three are used for a better signal-to-noise ratio). The twist angle is determined using $\lambda = \frac{a}{2 \sin(\frac{\theta}{2})}$, where $\lambda = \frac{1}{3} \sum_{j = 1}^3 \lambda_j$ and a is the lattice constant of graphene. Using the Fourier transform is generally more accurate than fitting three moir\'e lattice peaks, because it averages over the whole field of view, but this also limits its spatial resolution to the full field of view.\\
In this work, we introduce an alternative method of quantifying and visualizing the heterogeneity in open devices, with sub-moir\'e lattice cell resolution over length scales of hundreds of nanometers. We develop a spatial lock-in method that enables one to map, with sub-moir\'e wavelength resolution, the local twist angle $\theta^*(\vec{r})$, the local moir\'e anisotropy $\kappa(\vec{r})$, and the anisotropy direction $\psi(\vec{r})$, as defined below. Notably, we can separate these effects from each other and from rotations of the lattice (figure 1b). We then apply our method to determine the heterogeneity in open TBG devices.\\

\section{methods}
We fabricate our devices using the tear and stack method with a special focus on avoiding contamination to ensure the large clean areas needed for this study. A single graphene flake is pre-cut in halves with an AFM tip, ensuring initial crystallographic alignment between them. The first half is subsequently picked up with a hBN flake, mechanically exfoliated on a SiO$_2$/Si chip and adhered to a PDMS/PC stamp at $\sim100\ ^\circ$C. The second half of graphene is manually rotated to a target twist angle of 1.5$^\circ$--2.0$^\circ$ and consequently picked up by the hBN/graphene stack on PC. In the next step, the PC layer is carefully peeled off of the initial PDMS stamp and placed on another PDMS stamp up-side down. The sacrificial polycarbonate (PC) layer is then removed in 1-Methyl-2-Pyrrolidone. Subsequently, the TBG/hBN heterostructure is transferred on a target SiO$_2$/Si substrate with a pre-patterned navigation structure, two gold electrodes and a graphite gate contacting one of them within the measurement area. We carefully align the TBG/hBN stack with the local graphite gate to avoid short circuiting. The second pre-patterned gold electrode is used to electrically contact TBG using another graphite piece. The devices are inserted into our ultra-high-vacuum setup and annealed at 350$^\circ$C for 12h before inserting them into the low-temperature STM operating at 4.2K. The TBG samples are located using a capacitive navigation scheme \cite{n27}.\\

\begin{figure*}
\centering
\includegraphics[width=\linewidth]{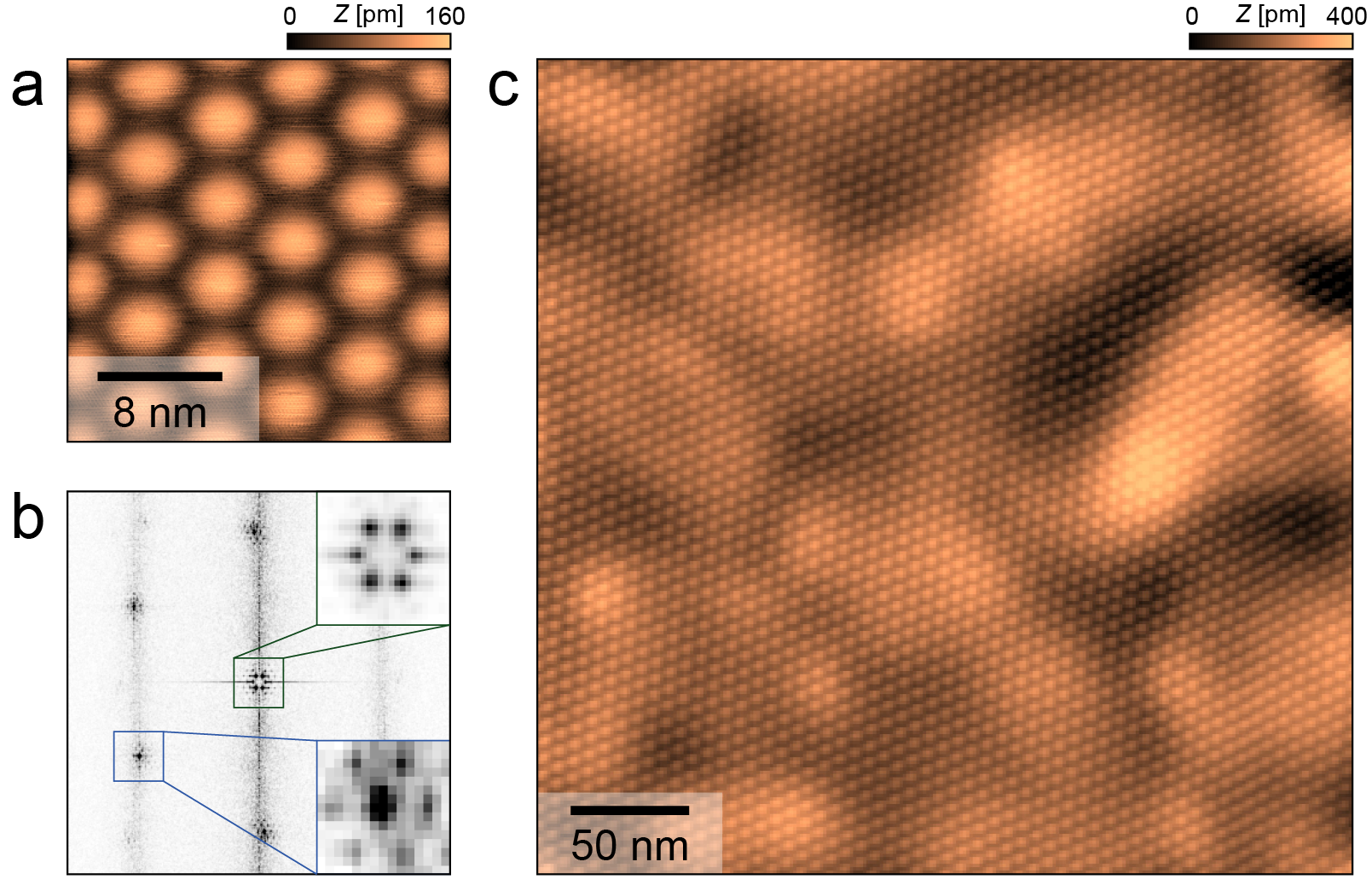}
\caption{a) STM topography of a device with an average twist angle of $\theta = 2.38^\circ$ (set-up conditions: V = 250 mV, I = 100 pA). The topography shows both the atomic- and moir\'e lattice. b) Fourier transform of a, with zoom ins of the moir\'e peaks (green inset) and the bottom left atomic peak (blue inset). Satellite peaks of the moir\'e lattice are visible around the atomic peak as well. c) Large scale topography measured on a different device with an average twist angle of 2.02$^\circ$ (set-up conditions: V = 250 mV, I = 20 pA).}
\end{figure*}

\section{Spatial lock-in algorithm}
Figure 2a shows a topographic image where both the atomic lattice of the top graphene layer and the moir\'e lattice are resolved. The Fourier transform of the image shows the lattice peaks as well as the peaks from the moir\'e superlattice (figure 2b, blue and green inset respectively).  While such small field-of-views are well suited for spectroscopy studies, we require large field of views that encompass many moir\'e cells for the heterogeneity study using spatial lock-in detection presented here. Figure 2c shows an example.

\begin{figure}
\centering
\includegraphics[width=\linewidth]{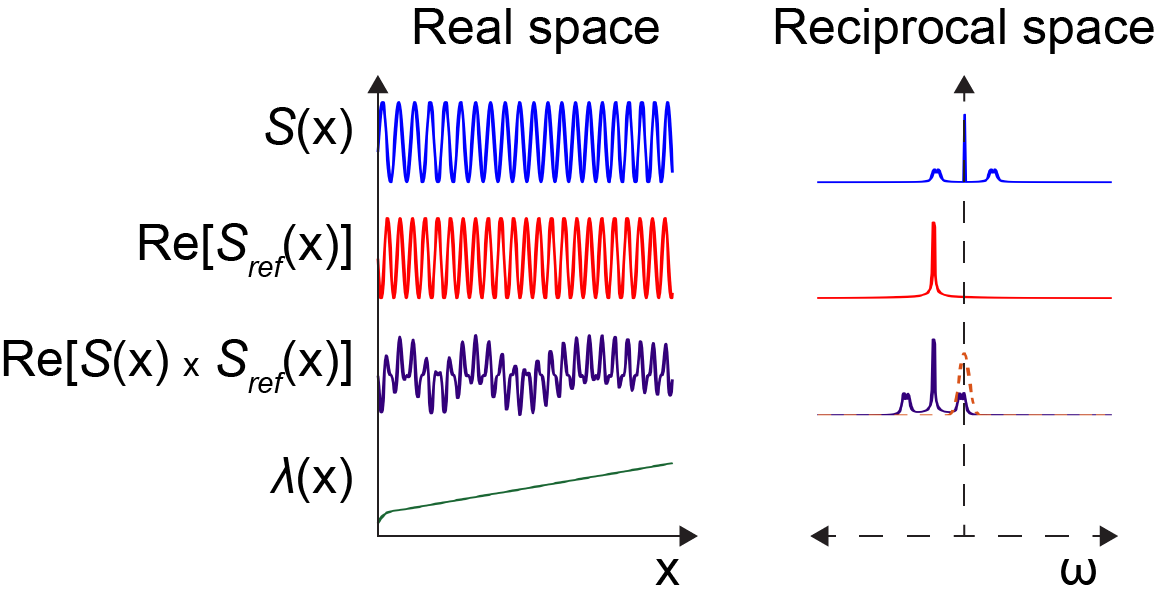}
\caption{Lock-in in 1D. The panels in the left column show, from top to bottom, the signal (an almost periodic sinusoid), the real part of the reference, the real part of the product of the signal and reference, and the wavelength calculated by taking the gradient of the phase of the product signal. The right column displays the Fourier transform of the (complex) signals in the left column. Finally, in the bottom right curve, the orange dashed line represents the gaussian filter used for the lock-in procedure.}
\end{figure}

The general method of spatial lock-in is illustrated in figure 3a for the one-dimensional case: the “measured” signal $S(x)$, a not-quite periodic signal, is multiplied with a reference signal, a perfectly periodic complex plane wave $S_\textit{ref}(x)$. The phase of the resulting signal, when low pass-filtered, corresponds to the local phase of the original wave. To obtain the local variations in wavelength $\lambda(x)$ of the original wave, shown on the bottom, one calculates the derivative of the local phase. Spatial lock-in algorithms like this have been used previously in electron microscopy studies (known as geometric phase analysis) \cite{n28,n29,n30} and optical metrology \cite{n31}. In the context of STM, the most well-known application is known as the Lawler-Fujita algorithm \cite{n32}. Lawler, Fujita et al. have, based on earlier work by Slezak et al. \cite{n33}, introduced a lock-in algorithm to correct topographic images for drift by calculating the displacement field, i.e. the vectors that connect the coordinates of the measured images with the points of an ideal reference lattice. Our motivation here is different: we do not need to correct an imperfect image, but want to extract heterogeneities of the lattice.\\
To do so, we start with defining three reference plane waves $R_j(\vec{r}) = e^{i \vec{q}_j \cdot \vec{r}},\ j \in \{1,2,3\}$, where the reference wavevectors $q_j$ are determined by simultaneously fitting six gaussians to the Bragg peaks in the Fourier transform of the topography (figure 4b). In order to measure deviations from an isotropic triangular lattice, we force the reference wavevectors to be of equal magnitude and 60 degrees with respect to each other (although see appendix A4 on choice of reference vectors). The reference lattice is then defined as the real part of the sum of the reference plane waves, i.e. $T_r(\vec{r}) = \text{Re}\left[ T_0 \sum_j R_j(\vec{r}) \right] = T_0 \sum_j \cos(\vec{q}_j \cdot \vec{r})$, where $T_0$ is the average amplitude.\\
The transformation between the measured lattice, $T_m(\vec{r})$, and this perfectly periodic, hexagonal reference lattice, $T_r(\vec{r})$ can approximately be parametrized as the shifts between points in the moir\'e lattice and corresponding points in the reference lattice. To this end, we introduce the displacement field, $\vec{u}(\vec{r})$, in the following manner:\\
\begin{equation*}
    T_m(\vec{r}) = T_r(\vec{r} + \vec{u}(\vec{r})) = T_0 \sum_j \cos\left(\vec{q}_j \cdot (\vec{r} + \vec{u}(\vec{r}))\right) .
\end{equation*}\\
To extract the displacement field from our data, each reference signal is multiplied with the original topographic image and low-pass filtered with a gaussian window. This operation corresponds to convolution of the original topographic image with the plane wave encompassed by a gaussian, calculating the relevant wave vector component of the ultimately small window 2D Fourier transform. The window of the gaussian filter needs to be chosen large enough (small enough in frequency space), in order to exclude larger frequencies, but simultaneously small enough (big enough in frequency space) to maintain good spatial resolution (appendix A). A lower limit on the filter size is put in place by higher frequencies: If the filter is chosen too small (too large in frequency space), more higher frequencies become included in the filter window reducing the signal to noise ratio. In practice a filter width of a few periods is used, as illustrated by the circle in figure 4a. The local phase of the result of this operation corresponds to the local shift between the real image and the reference wave, or more precisely $\phi_j(\vec{r}) = \vec{q}_j \cdot \vec{u}(\vec{r})$ (appendix A).\\
This local phase is $2\pi$ periodic and needs to be phase-unwrapped to remove discontinuities. After phase unwrapping, the displacement field $\vec{u}(\vec{r})$ can be extracted from two of the phase maps by pixel-wise multiplication with $Q^{-1}$, the inverse of a matrix containing the used wave vectors (Although not applied here, using all three wave vectors is more involved but can be beneficial for low signal-to-noise ratio situations, as detailed in appendix A4).\\ 
In a second step, we decompose the obtained displacement field, $\vec{u}(\vec{r})$ into the local effective twist angle, $\theta^*(\vec{r})$  and the local moir\'e anisotropy magnitude and direction, $\kappa(\vec{r})$ and $\psi(\vec{r})$ respectively. To that end, we consider the Jacobian of the transformation, 
$J = I + \vec{\nabla} \vec{u}$, which is the displacement gradient tensor that describes the transformation of an infinitesimal triangle at each position. The polar decomposition $J = WA$ splits $J$ into the product of the unitary matrix $W$, describing the local rotation of the lattice and a matrix $A$, describing the local scaling and anisotropy. This matrix $A$ can be further decomposed into a (unitary) rotation matrix $V$, indicating the major and minor axis of scaling and a diagonal scaling matrix $D$ such that $J = WA = WV^TDV$. This final decomposition is illustrated in figure 1b and makes it straightforward to extract relevant quantities. The change in density of unit cells is equal to the change in area under the effect of the deformation gradient tensor, hence the geometric mean of the scaling elements in the diagonal of $D$, $\sqrt{d_1 d_2} = \sqrt{\det(J)}$, allows us to calculate the wavelength of the moir\'e lattice and consequently, the local twist angle (appendix A). Furthermore, the local anisotropy magnitude, $\kappa(\vec{r})$, is calculated by taking the ratio of the scaling elements that make up $D$, $\kappa = \frac{d_1}{d_2}$, where $d_1 > d_2$. Defined in this way, $\kappa = 1$ indicates an isotropic lattice, and  $\kappa > 1$ indicates an anisotropy of the moir\'e lattice in the direction given by $\psi$, the angle corresponding to the rotation corresponding to $V$. Lastly, the rotation of the total lattice, corresponding to $W$, is left unattended, as a rotation of the full lattice should not directly influence the physics at play, although we point out that it does describe the variations of the rotation with respect to the hBN substrate.\\
\vspace{-20pt}

\section{Results}
Figure 4c shows the effective twist angle $\theta^*(\vec{r})$, figure 4d the local anisotropy $\kappa(\vec{r})$ and figure 4e shows the angle of major scaling $\psi(\vec{r})$, all as a function of location for open-device TBG. The maps show rather smooth variations with an exception in the bottom right corner of the field of view, where an apparent vertical feature appears. This feature is only barely visible in the topography itself, showcasing the sensitivity of our method. The origin of this particular vertical stripe remains unclear, and no such peculiarities were observed in our other data (appendix G).
\\
The overall twist angle heterogeneity in the image in figure 4c, excluding border effects, is 0.033$^\circ$ (standard deviation) or 0.23$^\circ$ peak-to-peak. We find areas of hundreds of nanometers with a standard deviation of the twist angle of 0.02$^\circ$ and a peak to peak variation of 0.08$^\circ$, e.g. in the area marked by a red square in figure 4c. 
A good estimate of the accuracy of our method can be made by applying the conventional Lawler-Fujita algorithm \cite{n32} and using spatial lock-in to extract the residual displacement field (appendix E). We find residual twist angle variations more than one order of magnitude smaller than the originally found values, underlining the accuracy of our method. We further note that this is achieved with a pixel density corresponding to $\sim$5 pixels per moir\'e lattice constant, which makes implementation of the conventional heterostrain model challenging (appendix J).\\
Our result allows for a first comparison between open and encapsulated devices. For the latter, we compare our results with results from SOT \cite{n14}. SOT measures the superlattice density, $n_s(\vec{r})$, which scales directly with the size of the unit cell. 
We note that SOT does not differentiate between heterogeneity of the chemical potential, strain, and twist-angle, which can all influence $n_s(\vec{r})$.\\
To make a  comparison between SOT and our data, one has to take into account the difference in the width of the point spread function (PSF). As this width is around $30$ nm for SOT, we artificially broaden the PSF of our data to match (appendix I), which naturally leads to a reduction of both the peak-to-peak spread and the standard deviation. In the full field of view, including the bright vertical feature, we find a peak to peak spread of 0.20$^\circ$ and a standard deviation of 0.036$^\circ$. These numbers are similar among different devices of similar twist angle (appendix K) and measured for areas of several hundreds of nanometers across.

Interestingly, this result matches rather well with the result from SOT on encapsulated devices, despite the lack of a stabilizing top hBN slab in our devices. This implies that open devices can rival the quality of encapsulated devices, at least in terms of twist angle homogeneity.

Our results then raise the following question: why have open-devices never been shown to superconduct, nor to show spectral gaps in low temperature tunneling experiments? Assuming that the mechanical properties of the bilayer do not change drastically as the angle of reconstruction is approached ($\theta \approx 1.0^\circ$), our experiments suggest that the homogeneity of the TBG itself cannot be the only reason. Instead, another reason might be the absence of a second hBN layer encapsulating the bilayer, despite hBN often being neglected in theoretical studies due to its supposed weak interaction. Furthermore, the second hBN layer creates a near symmetric environment for the bilayer. We speculate that breaking of this symmetry may be at the basis for the lack of superconductivity in open devices. However, more careful transport investigations of open devices are necessary to confirm this hypothesis.\\

\begin{figure*}
\centering
\includegraphics[width=\linewidth]{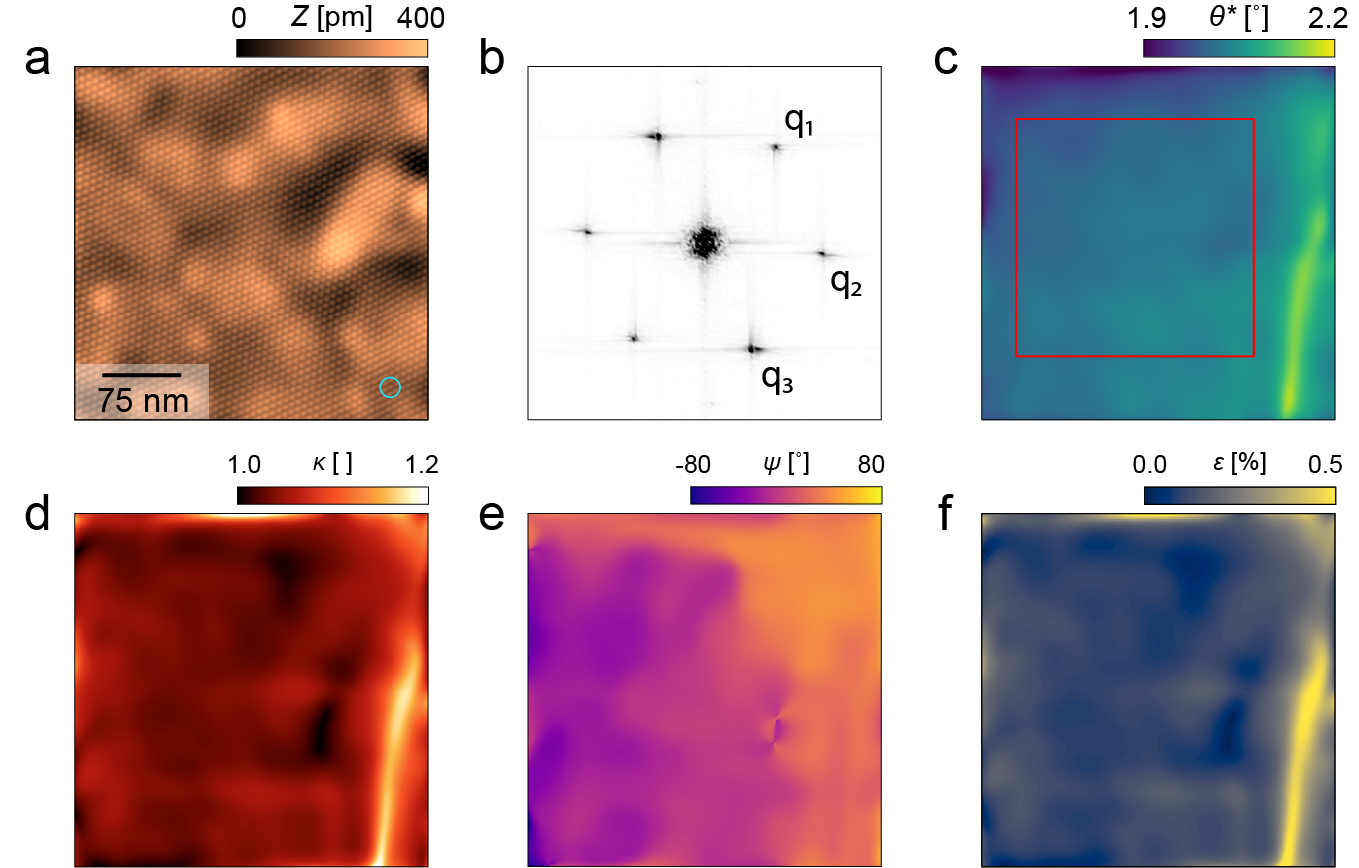}
\caption{a) STM topography of a device with an average twist angle of 2.02$^\circ$ (V = 250 mV, I = 20 pA, same data as Figure 2c). The blue circle in the bottom right indicates the size of the filter used by the algorithm (see main text). b) Fourier transform of a, showing the Bragg peaks of the moir\'e lattice visible in the image. The Bragg peaks are labelled $q_1 - q_3$. c) Effective twist angle map extracted from a, by the algorithm discussed. The red square indicates the area over which the average twist angle and standard deviation are calculated. d) Local moir\'e anisotropy map $\kappa(\vec{r})$ extracted by the algorithm from a. e) Local moir\'e anisotropy direction $\psi(\vec{r})$ extracted by the algorithm from a. f) Heterostrain map extracted as described in the text.}
\end{figure*}

The local anisotropy parameter $\kappa(\vec{r})$ discussed here can be related to heterostrain, following the model of Kerelsky et al. \cite{n22}. Here, it is assumed that one of the graphene sheets is strained with a uniaxial strain $\epsilon(\vec{r})$, while the other one is unaffected and only undergoes a rotation. To connect to our measurements, we note that for small average twist angles, the displacement field of the moir\'e lattice is related to relative displacement of the constituting layers by the following formula: $(\langle J\rangle - I) \cdot \vec{u}_\textit{moir\'e}(\vec{r}) = \vec{u}_\downarrow(\vec{r}) - \vec{u}_\uparrow(\vec{r}) = \vec{u}_\sim(\vec{r})$, where $\langle J\rangle$ is the Jacobian corresponding to the average angle between the layers and $\vec{u}_\sim(\vec{r})$ is the relative displacement field experienced between the two sheets (appendix B). The relative displacement field can be decomposed in the same way as before, where the angle corresponding to $W$ now corresponds to the deviation of the twist angle between the two sheets from the average twist angle, and the local anisotropy $\kappa(\vec{r})$ and $\psi(\vec{r})$ obtained from the resulting scaling matrix indicate the relative strain between the layers. Furthermore, from the resulting scaling matrix elements, we can calculate the magnitude of the strain applied to the deformed sheet, $\epsilon(\vec{r})$ (appendix B). We show the resulting $\epsilon(\vec{r})$ in figure 4f. On average, we find that $\epsilon = 0.14\%$ with a standard deviation of 0.09\%.\\
It is interesting to compare the numbers for strain and twist angle heterogeneity, and their respective influence on the electronic structure of TBG. Calculations using a continuum model have considered both strain and twist angle changes in TBG samples close to the magic angle \cite{n34}. It was shown that a heterostrain of $\epsilon \approx 0.1\%$ results in a splitting of the van Hove singularities of approximately 5 meV. This is comparable to variations in the twist angle of about 0.03$^\circ$, which we obtain by interpolating the relation between twist angle and van Hove splitting given in \cite{n35}. Furthermore, stress can cause strong qualitative changes to the electronic structure including new van Hove singularities for $\epsilon \approx 0.5\%$. If we compare these numbers with our measurements, we conclude a roughly equal effect of the observed strain and twist angle inhomogeneity, suggesting that both have to be taken into account when fabricating samples, as both effects significantly alter the electronic structure compared to a perfect lattice.\\ 
Before concluding, we want to address one potential challenge of the method introduced here: it is also sensitive to piezo drift. Piezo drift occurs in STM experiments due to thermal fluctuations that influence the piezo, due to piezo relaxation after a change of field of view, or due to the piezo relaxation from the movement necessary to take the topography. The former two effects change over time. The latter effect depends on the speed with which the topography is measured. To check the validity of this procedure, we have repeated the above procedures for different topographies in the same field of view, taken with different scan speeds at different times. As we show in detail in the SI, these different measurements yield very similar results, demonstrating that the twist angle variations we measure are intrinsic and not a consequence of piezo drift.\\

\section{Conclusion}
In this work, we have visualized and characterized structural heterogeneity in TBG, demonstrating peak to peak variations in the twist angle of roughly 0.08$^\circ$ over areas of hundreds of nanometers. While our samples exhibit an average twist angle higher than the magic angle, we expect the issues to be similar as long as the twist angle is above the reconstruction that occurs for twist angles $\lesssim 1^\circ$. This indicates that the best open device TBG could, purely based on homogeneity of the twist angle, superconduct, and that lack of experimental evidence thereof suggests a critical role of the missing hBN top layer. The spatial lock-in algorithm we introduced is in principle applicable to a variety of different moir\'e materials, and additionally, may also be usable in a different context, e.g. in determining the topological properties of band structures through QPI measurements \cite{n38}. We anticipate that this algorithm can be applied to other microscopy probes as well, including AFM and LEEM. Lastly, by presenting our results in the way we did, we hope to pave the way for further studies, especially for correlating electronic- and spatial properties by combining with theoretical models like the ones presented in references \cite{n34,n36,n37}.
\vspace{-15pt}

\begin{acknowledgments}
We thank H. Zandvliet, P. Koenraad, P. Neu and R. Wijgman for valuable discussions. Furthermore, we thank K. van Oosten, F. Groenewoud, D. Scholma and T. Mechielsen for technical support. This work was supported by the European Research Council (ERC StG SpinMelt) and by the Dutch Research Council (NWO), as part of the Frontiers of Nanoscience programme, as well as through a Vidi grant (680-47-536). D.K.E. acknowledges support from the Ministry of Economy and Competitiveness of Spain through the “Severo Ochoa” program for Centres of Excellence in R\&D (SE5-0522), Funda-ci\'o  Privada Cellex, Fundaci\'o Privada Mir-Puig, the Generalitat de Catalunya through the CERCA program, the H2020 Programme under grant agreement n$^o$ 820378, Project: 2D·SIPC and the La Caixa Foundation.
\end{acknowledgments}

%

\appendix
\section{Spatial Lock-in Algorithm}
\subsection{Deformations of a lattice}\label{sec:deformations}
We perform lock-in measurements on images that clearly display a periodic lattice.
In STM, this implies we can use any topography of sufficient quality that displays the crystal lattice.
The idea is to use a lock-in measurement in order to find a transformation of coordinates between the measured, “distorted” image and its pristine, undeformed equivalent (in this work, a perfect triangular lattice). Defining the measured and pristine image as
$T_m(\vec r), T_r(\vec r')$ respectively, both with measurement coordinates $\vec r = (x,y) \in \mathbb{R}^2$ and lattice coordinates $\vec r' = (x',y') \in \mathbb{R}^2$, the following relation holds:

$$T_m(\vec r) = T_r\left(\vec r + \vec u (\vec r)\right) = T_r\left(\vec f(\vec r)\right) = T_r( \vec r')= T_m\left(\vec f^{-1}(\vec r')\right)$$
where the transformation from measurement coordinates to lattice coordinates is given by:  
\begin{equation}\label{eq:u_definition}
\vec f(\vec r) = \vec r + \vec u (\vec r) = \vec r'    
\end{equation}

Here, $\vec u (\vec r)$ is called the displacement field, connecting the measurement coordinates to the lattice coordinates, as is well-established in continuum mechanics. 
For convenience, we also define the inverse displacement:
$$\vec u'(\vec r') := \vec f^{-1}(\vec r') - \vec r' = \vec r - \vec r'$$
Note that by substitution, we have the following relation between forward and inverse displacement:
$$\vec u'(\vec r') = \vec f^{-1}(\vec f(\vec r)) - (\vec r + \vec u(\vec r)) = -\vec u(\vec r)$$
With this, we can express the pristine image at lattice coordinates in terms of the measured image:
\begin{align*}
T_r(\vec r') &= T_m\left(\vec f^{-1}(\vec r')\right) = T_m(\vec r' + \vec u'(\vec r'))\\
&= T_m\left(\vec r' - \vec u(\vec r)\right) \\
&= T_m\left(\vec r' - \vec u(\vec r'- \vec u(\vec r))\right) \\
&\approx T_m\left(\vec r' - (\vec u(\vec r') - (\nabla \vec u) (\vec r' - \vec r))\right)\\
&=T_m\left(\vec r' - (\vec u(\vec r') + (\nabla \vec u) \vec u(\vec r))\right)\\
&=T_m\left(\vec r' - \vec u(\vec r') + (\nabla \vec u) \vec u'(\vec r'))\right)\\
\end{align*}

Therefore, if we can determine $\vec u(\vec r)$, and thereby $\vec u'(\vec r')$, we can reconstruct the  pristine image. This is the idea of the Lawler--Fujita reconstruction algorithm~\cite{n32}. In their original paper, Lawler--Fujita uses $\vec u'(\vec r') = -\vec u (\vec r')$, which is a good approximation if $\vec u$ varies slowly.

\subsection{Properties of the deformation}
The displacement field $\vec u (\vec r)$ as defined above, fully describes the deformation of the lattice, but does not directly provide insight into the relevant properties. To that end, we first define the Jacobian of the transformation $\vec f$: 
$$J \equiv \nabla \vec f = \mathbbm{1} + \nabla \vec u$$, where $\nabla \vec u$ is the Jacobian of the displacement field, in continuum mechanics terminology the deformation gradient tensor, and in canonical terms defined as follows: 
$$
\vec \nabla \vec u = 
\begin{pmatrix}
\frac{du_x}{dx} & \frac{du_x}{dy}\\
\frac{du_y}{dx} & \frac{du_y}{dy}\\
\end{pmatrix}$$

In order to fully characterise the deformation of the lattice, we decompose $J$ in its polar form: 
\begin{equation}
J = WP = WV^\top DV\ ,
\label{eq:decomposition}
\end{equation} 
where $W$ is the rotation matrix corresponding to the rotation of the full lattice and the matrix $P$ describes the local anisotropy and scaling. 
$P$ is further decomposed in the rotation matrix $V$ indicating the orientation of the axis of anisotropy (i.e. the axis of largest scaling, with the axis of smallest scaling perpendicular to it) and the diagonal scaling matrix 
$D= 
\begin{pmatrix}
d_1 & 0\\ 
0 & d_2\\ 
\end{pmatrix}$, 
where by convention and implementation $d_1 \geq d_2$ holds for any position $\vec r$.

The geometric mean of these directional scaling factors is equal to the square root of the determinant of $D$ and therefore of $J$: $\sqrt{d_1d_2} = \sqrt{\text{det}(J)}$.
As this corresponds to the local scaling of the wavelength of the moir\'e lattice, we can use this to quantify the local twist angle:

\begin{equation}
    \lambda (\vec r) = \sqrt{d_1d_2}\frac{4\pi}{\sqrt{3} |\vec q_j|}
\end{equation}
Where $|\vec q_j|$ is the length of the chosen reference vectors.
This local wavelength is then converted to a local twist angle using the well-known expression:
$$\theta(\vec r) = 2 \arcsin\left(\frac{2\lambda(\vec r)}{a}\right)$$,
where $a=2.46\text{\AA}$ is the lattice constant of graphene and $\theta(\vec r)$ the local twist angle.

A quantification of the local anisotropy is given by the ratio $\kappa = d_1/d_2$ and the angle between the anisotropy axis and the $x$-axis is finally calculated from $V$: $\psi = \arctan \left(\frac{V_{xy}}{V_{xx}}\right)$. 

In our practical implementation, the singular value decomposition (SVD) is used to obtain the decomposition in equation \ref{eq:decomposition} for each point in the image, and Matlab's \texttt{atan2} is used to find the right quadrant of the angles from the signs of $V_{xx}$ and $V_{xy}$.

\subsection{Determination of the displacement field $\vec u(\vec r)$}
In order to determine $\vec u(\vec r)$ for a certain image, we perform a lock-in measurement. To clarify, we can represent any (nearly) periodic image as:
\begin{equation}
    T_m(\vec r) = T_0\sum_j e^{i\vec q_j\cdot \left(\vec r+\vec u (\vec r)\right)} = 
T_0\sum_j e^{i(\vec q_j\cdot \vec r + \phi_j)}
\end{equation}
where $\phi_j = \vec q_j\cdot \vec u(\vec r)$ is the position-dependent phase of the lattice. The summation runs over the reciprocal lattice vectors $\vec q_j$ ($j \in \{1,2,3\}$ for a hexagonal lattice),
$T_0$ is the constant indicating the amplitude of the modulation and $\vec u (\vec r)$ is again the displacement field.

The phase is measured using standard lock-in procedure: 
The existing image is mixed with a reference image containing a specific plane wave. 
If we choose the periodicity of this reference wave sufficiently close to that of the lattice in the image itself, we can then low-pass filter the mixed image and end up with a phase map for a specific wave. For clarification:

$$\cos(\vec q_j\cdot \vec r + \phi_j) e^{-iq_j\cdot \vec r}
= \frac{e^{i \phi_j }}{2}\left( 1+ e^{-2i(\vec q_j\cdot \vec r + \phi_j)}\right)
\mapsto \frac12 e^{i \phi_j }$$
where the cosine in the first term denotes the (real-valued) measured image, whereas the complex exponential denotes the reference wave and $\mapsto$ denotes low-pass filtering in order to get rid of the last term between brackets, corresponding to a rotating wave approximation. Alternatively, for a gaussian low-pass filter, this corresponds to a real space gaussian integration window of the lock-in.

By taking the (pointwise) angle of the complex, filtered product image, we end up with the phase map. 
In particular, this phase map contains information about the displacements of each pixel in the measured image $T_m(\vec r)$ with respect to the pristine reference lattice $T_r(\vec r)$ along the wave vector $\vec q_j$ used for the lock-in procedure.
This procedure is repeated for at least one additional reciprocal lattice vector. The two phase maps are then used to find the displacement field $\vec u (\vec r)$.
From the definition of $\vec u(\vec r)$ (eq. \ref{eq:u_definition}), the following holds: $\vec r' = \vec r + \vec u(\vec r)$. Multiplying this equation by the reciprocal lattice vectors, we get a system of equations expressing the projection of the distortion onto the reciprocal lattice vectors:
$$\vec q_j \cdot \vec r' = \vec q_j \cdot \vec r + \phi_j,\ j\in {1,2,3}$$

Selecting only $j\in\{1,2\}$, we have in matrix notation:
$$Q = 
\begin{pmatrix}
-\vec q_1 -\\
-\vec q_2 -\\
\end{pmatrix}
=
\begin{pmatrix}
q_{1x} & q_{1y}\\
q_{2x} & q_{2y}\\
\end{pmatrix}
$$ such that we can write for $\vec \phi = \binom{\phi_1}{\phi_2}$:
\begin{equation}
Q\vec r' = Q\vec r + \vec \phi\ .
\label{eq:linearQsystem}
\end{equation}
Multiplying by $Q^{-1}$, we find $\vec r' = \vec r+ Q^{-1}\phi$, and therefore $\vec u (\vec r) = Q^{-1}\vec \phi(\vec r)$.

\subsection{Additional notes on choice of reference vectors}
\subsubsection{Selecting two reference vectors}
To obtain $\vec u(\vec r)$ as described above, we only used the phase of the lock-in signal of two reference vectors. For a triangular/hexagonal lattice, \textit{a priori} three possible choices of which two reference vectors to use are possible from the three linear independent references vectors as fitted to the FFT of the image. To select which two vectors to use for the reconstruction of $\vec u(\vec r)$, we either selected the ones with the largest average lock-in amplitude, or by inspecting the phase-unwrapped images and selecting the ones where no remaining phase slips occurred.
\subsubsection{Using more than two reference vectors}
In principle, information is lost when only selecting the phase of the lock-in signal of two reference vectors to obtain $\vec u(\vec r)$. Although not used in this work, in low signal-to-noise ratio situations, it could be beneficial to use all the information. Equation \ref{eq:linearQsystem} also holds for more than two phases and reference vectors. Although $Q$ is not a square matrix in this case, a solution can be obtained for each pixel using linear least squares minimization of the following equivalent equation:
$$
Q \vec u(\vec r) = \phi(\vec r)
$$
Where additionally the amplitude of the lock-in signals can be used as weights to the minimization problem.
\subsubsection{Isotropy}
Enforcing the reference lattice to be isotropic can be done either in advance, by enforcing isotropic reference wavevectors (as applied in this work) or alternatively, after the initial lock-in step, by adding an additional linear phase $\Delta \phi_j = \vec{\Delta q}_j \cdot \vec r$  to the obtained phase, where $\vec{\Delta q}_j$ is the difference between the used reference wavevector and the isotropic wavevector. 

The advantage of the latter method would be a slightly improved signal-to-noise ratio, as the smoothing window can be centered around the actual average wavevector occurring in the image instead of the ideal, equal-length, 60 degree rotated ones.

\section{Relation of moir\'e lattice to relative displacement}

For a non-homogeneous bilayer, the system can be fully described by two displacement fields $\vec u_\uparrow(\vec r), \vec u_\downarrow(\vec r)$ of respectively the top and bottom layer compared to an undistorted system.

\begin{align*}
T_m(\vec r) &= T_{m\uparrow}(\vec r) \oplus T_{m\downarrow}(\vec r)\\ &=
T_{r\uparrow}\left(\vec r+\vec u_\uparrow(\vec r)\right) \oplus T_{r\downarrow}\left(\vec r+ \vec u_\downarrow(\vec r)\right)
\\&= T_{r\uparrow}\left(\vec r_{\uparrow}\right) \oplus T_{r\downarrow}\left(\vec r_{\downarrow}\right)
\end{align*}

where $T_r(\vec r)$ denote the atomic lattices, $\vec r_{\downarrow}, \vec r_{\uparrow}$the lattice coordinates of both lattices and $\oplus$ denotes the (as of now, unspecified) operation of the combination of two lattices into one image.

We can express the deformation of one atomic lattice w.r.t the coordinates of the other:
\begin{align*}
T_{r\downarrow}\left(\vec r_{\downarrow}\right) &= T_{r\downarrow}\left(\vec r+\vec u_\downarrow(\vec r)\right) = 
T_{r\downarrow}\left(\vec f_{\downarrow}(\vec r)\right)\\ &= 
T_{r\downarrow}\left(\vec f_{\downarrow}(\vec f^{-1}_\uparrow(\vec r_{\uparrow}))\right)
= T_{r\downarrow}\left(\vec f_{\downarrow}(\vec r_{\uparrow}+ \vec u'_\uparrow(\vec r_{\uparrow}))\right)\\
&= T_{r\downarrow}\left(\vec r_{\uparrow}+ \vec u'_\uparrow(\vec r_{\uparrow})+\vec u_\downarrow(\vec r_{\uparrow}+ \vec u'_\uparrow(\vec r_{\uparrow}))\right)
\end{align*}

Assert $\vec  u_\downarrow(\vec r) = J_\downarrow \vec r + \vec v_\downarrow(\vec r)$, i.e. a global rotation and/or scaling plus local variations. Note that here, $J_\downarrow$ is constant 2-by-2 matrix corresponding to a mean $\nabla \vec u$, and therefore corresponding to $J-I$ in terms of the $J$ defined in the previous section. In this case, we have:
$$T_{r\downarrow}\left(\vec r_{\downarrow}\right)  = T_{r\downarrow}\left((I+J_\downarrow)(\vec r_{\uparrow}+ \vec u'_\uparrow(\vec r_{\uparrow}))+\vec v_\downarrow(\vec r_{\uparrow}+ \vec u'_\uparrow(\vec r_{\uparrow}))\right) $$

For two real lattice plane waves $T_r(\vec r') = \cos(\vec q_j \cdot \vec r')$ and taking pointwise product for the $\oplus$ operator, we have:

\begin{align*}
T_m(\vec r_\uparrow) &= \cos(\vec q_j \vec r_{\uparrow}) \cdot \\
&\cos(\vec q_j \left[(I+J_\downarrow)(\vec r_{\uparrow}+ \vec u'_\uparrow(\vec r_{\uparrow}))+\vec v_\downarrow(\vec r_{\uparrow}+ \vec u'_\uparrow(\vec r_{\uparrow}))\right])\\
&= \cos(\vec q_j \vec r_{\uparrow})\cos(\vec q_j \vec r_{\uparrow} + \delta(\vec r))\\
&= \frac12\cos(2\vec q_j \vec r_{\uparrow}+\delta(\vec r))+\frac12 \cos(- \delta(\vec r))\\
&= \frac12\cos(2\vec q_j \vec r_{\uparrow}+\delta(\vec r))+\frac12 \cos(+ \delta(\vec r))\\
\intertext{For the modulation $\delta(\vec r)$ the following holds:}
\delta(\vec r) &= \vec q_j \left[J_\downarrow(\vec r_{\uparrow}+ \vec u'_\uparrow(\vec r_{\uparrow}))+\vec u'_\uparrow(\vec r_{\uparrow})+\vec v_\downarrow(\vec r_{\uparrow}+ \vec u'_\uparrow(\vec r_{\uparrow}))\right]\\
&=\vec q_j J_\downarrow\left[\vec r_{\uparrow}+ \vec u'_\uparrow(\vec r_{\uparrow}) + J_\downarrow^{-1}\left(\vec u'_\uparrow(\vec r_{\uparrow})+\vec v_\downarrow(\vec r_{\uparrow}+ \vec u'_\uparrow(\vec r_{\uparrow}))\right)\right]\\
\intertext{Substituting $\vec r_{\uparrow} = \vec r+\vec u_\uparrow(\vec r)$ and $\vec u'_\uparrow(\vec r_{\uparrow}) = -\vec u_\uparrow(\vec r)$:}
\delta(\vec r) &= \vec q_j J_\downarrow \left[\vec r- J_\downarrow^{-1}\vec u_\uparrow(\vec r)+J_\downarrow^{-1}\vec v_\downarrow(\vec r)\right]\\
&= \vec q_j J_\downarrow \left[\vec r+\vec u_\textit{moir\'e}(\vec r)\right]\\
\intertext{With $\vec u_\textit{moir\'e}(\vec r) = J_\downarrow^{-1}(\vec v_\downarrow(\vec r) - \vec u_\uparrow(\vec r)) = J_\downarrow^{-1}\vec u_\sim (\vec r)$, where $\vec u_\sim (\vec r)$ denotes the relative displacement between the upper layer and the rotated lower layer. Substituting back in $T_m$:}
T_m(\vec r)&= \frac12\cos(2\vec q_j (\vec r+\vec u_\uparrow(\vec r) +\frac12\left[J_\downarrow \vec r- \vec u_\uparrow(\vec r)+\vec v_\downarrow(\vec r)\right]))\\ &+ \frac12 \cos(J_\downarrow^\top \vec q_j  \left[\vec r+\vec u_\textit{moir\'e}(\vec r)\right])\\
T_m(\vec r)&= \frac12\cos(2\vec q_j (\vec r +\frac12\left[J_\downarrow \vec r+ \vec u_\uparrow(\vec r)+\vec v_\downarrow(\vec r)\right])) \\
&+\frac12 \cos(J_\downarrow^\top \vec q_j  \left[\vec r+\vec u_\textit{moir\'e}(\vec r)\right])\\
\end{align*}

Note that for a 2D lattice consisting of the sum of 2 or more cosines, each with it's own $\vec q_j$, this construction can be made for each $\vec q_j$ separately, nevertheless resulting in a single, joint $\vec u_\textit{moir\'e}(\vec r)$ (as expected).

For a small twist angle $\theta$ between two equal lattices, e.g. magic angle twisted bilayer graphene, we have:
\begin{align*}
J_\downarrow &=  R(\theta) - I
= \begin{pmatrix}
\cos\theta -1 &-\sin\theta\\
\sin\theta & \cos\theta-1 \\
\end{pmatrix}\\
&\approx \begin{pmatrix}
-\frac12\theta^2 +\frac{\theta^4}{24} &-\theta+ \frac{\theta^3}{6}\\
\theta - \frac{\theta^3}{6}& -\frac12\theta^2 +\frac{\theta^4}{24}\\
\end{pmatrix}\\
&= \theta \begin{pmatrix}
-\frac12\theta +\frac{\theta^3}{24} &-\left(1- \frac{\theta^2}{6}\right)\\
1 - \frac{\theta^2}{6}& -\frac12\theta +\frac{\theta^3}{24}\\
\end{pmatrix}\\
&= \theta R\left(\tfrac\pi 2 + \tfrac{\theta}{2}\right)+ 
\theta^3 \begin{pmatrix}
\frac{\theta}{48} &- \frac{1}{3}\\
 + \frac{1}{3}& \frac{\theta}{48}\\
\end{pmatrix}
\end{align*}
Therefore, in this case the topography $T_m(\vec r)$ consists of a sum of a cosine with approximately twice the atomic frequency and a cosine with approximately $\theta$ times the atomic frequency: the moir\'e frequency. As expected, this lattice is rotated by $\tfrac{\pi}{2}$ plus half the angle of the original rotation, i.e. angled halfway in-between both atomic lattices.

\subsection{Relation to uniaxial strain models}
Graphene has a Poisson ratio $\delta = 0.17$, so if a strain $\epsilon$ is applied in one direction, it shrinks in the perpendicular direction by $\delta \epsilon$. By applying the decomposition into $\theta(\vec r),\ \kappa(\vec r)$ and $\psi(\vec r)$ as described in Section \ref{sec:deformations}, to the relative displacement between the layers $\vec u_\sim (\vec r)$ and assuming the relative strain is dominated by the strain of one layer, we can calculate that strain $\epsilon(\vec r)$. For uniaxial strain, we have with these assumptions in terms of the decomposition into relative displacement:
\[\kappa(\vec r) = \frac{d_1}{d_2} = \frac{1+\epsilon}{1-\delta\epsilon}\]
and therefore we can express the strain of a single layer as follows:
\[\epsilon(\vec r) = \frac{d_1-d_2}{d_2+\delta d_1}\]
which can then be related to other measurements and models \cite{n34,n22}. In appendix J, we discuss the accuracy of these models.
Note that the measured quantity $\vec u_\textit{moir\'e}(\vec r)$ is related to the relative displacement by a multiplication of $J^{-1}$. For small twist angles, $\|J^{-1}\|\approx \frac{1}{\theta}$ (with $\theta$ in radians, i.e. for $\theta = 1.05^\circ$ we have $\|J^{-1}\|\approx 55$), strongly amplifying effects of small relative displacement.\\

\section{Phase unwrapping \& singularities}

\begin{figure}[!ht]
    \centering
    \includegraphics[width=\linewidth]{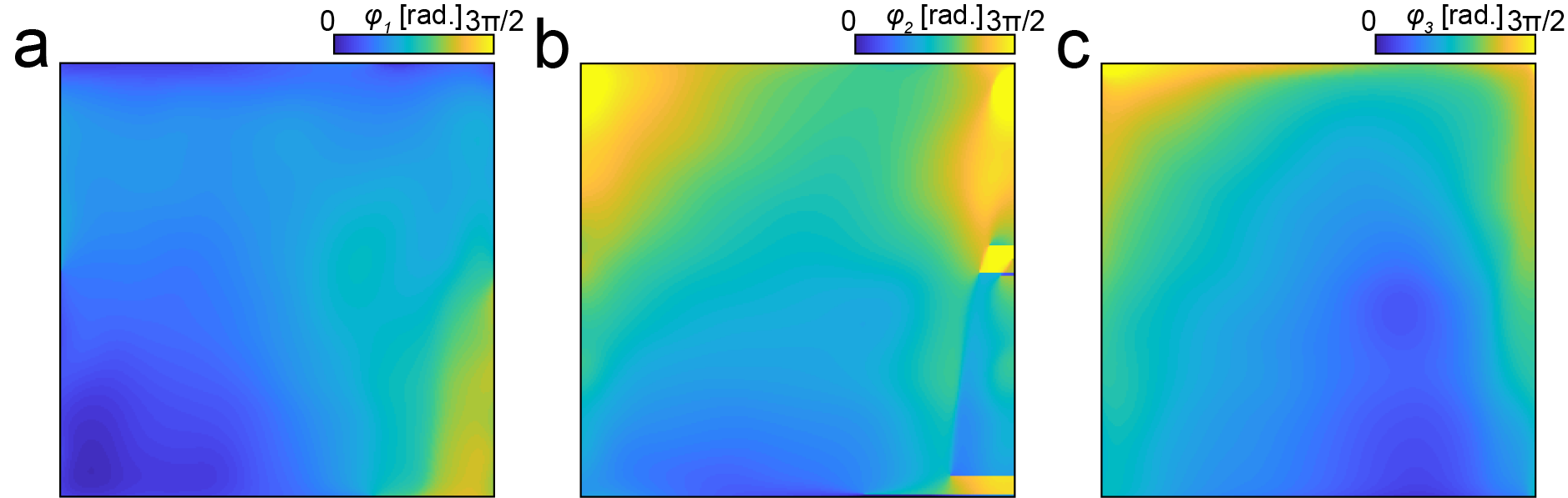}
    \caption{Phase maps of the data shown in main text figure 4, figure \ref{S2}a. a,b,c correspond to the phase maps of the Bragg peak labeled $\vec q_1$,$\vec q_2$ and $\vec q_3$ respectively (see main text figure 2b). Because the map corresponding to $\vec q_2$ shows some phase singularities, we use $\phi_1$ and $\phi_2$ for determining the displacement field.}
    \label{S1}
\end{figure}

In this work, phase unwrapping of a periodic phase is needed in two separate places: unwrapping the lock-in phase $\phi_j(\vec r)$ before reconstructing $\vec u(\vec r)$, and to obtain a single valued anisotropy angle $\psi(\vec r)$.
The phase is unwrapped in both directions of the image. 
The order in which this is done usually does not matter, provided there are no phase singularities present in the image. 
We occasionally encountered some phase singularities in one of the three phase maps (figure \ref{S1}), but we worked around this simply by using the other two phase maps in order to find the displacement field.\\

In case this is not an option, for example when applying this technique to a square lattice, and/or when phase slips are present in all phase maps, there are more sophisticated algorithms for phase unwrapping available: ~\cite{Ghiglia1994,Herrez2002,Kemao2007}.\\
Some of these phaseslips were present in the $\psi(\vec r)$ maps, for example the one displayed in the main text, figure 4e. Here, we used a Matlab implementation of a least-squares based phase unwrapping algorithm ~\cite{Ghiglia1994, Firman2016}.

\section{Device overview}
\begin{figure}[ht]
    \centering
    \includegraphics[width=1.2\linewidth]{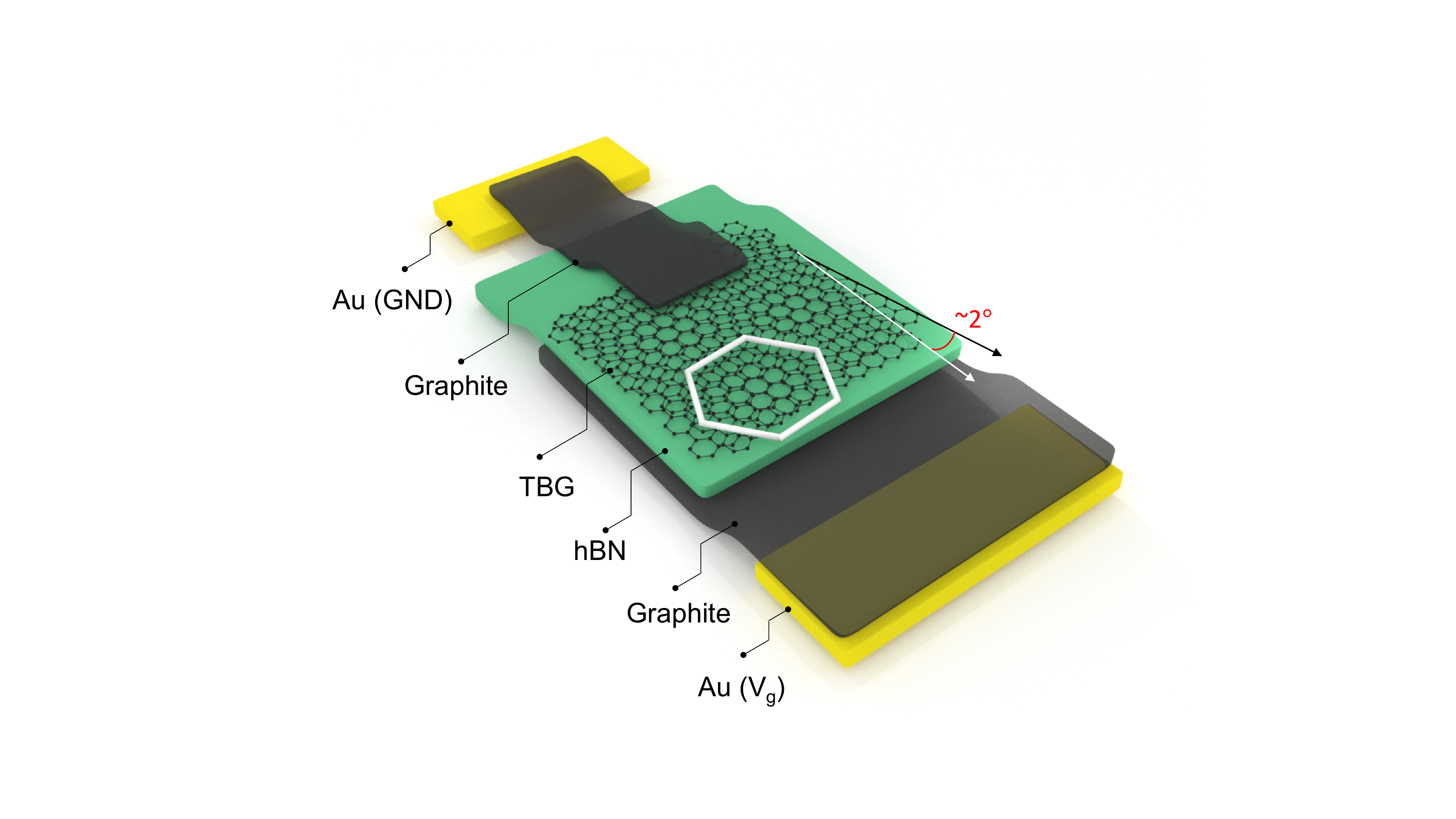}
    \caption{Schematic overview of the devices studied in this work.}
    \label{Sgeom}
\end{figure}
A schematic of the devices studied in this work is presented in figure \ref{Sgeom}. More information about the actual fabrication process can be found in the main text.

\section{Accuracy of the algorithm}
\begin{figure}[!hb]
\centering
\includegraphics[width=\linewidth]{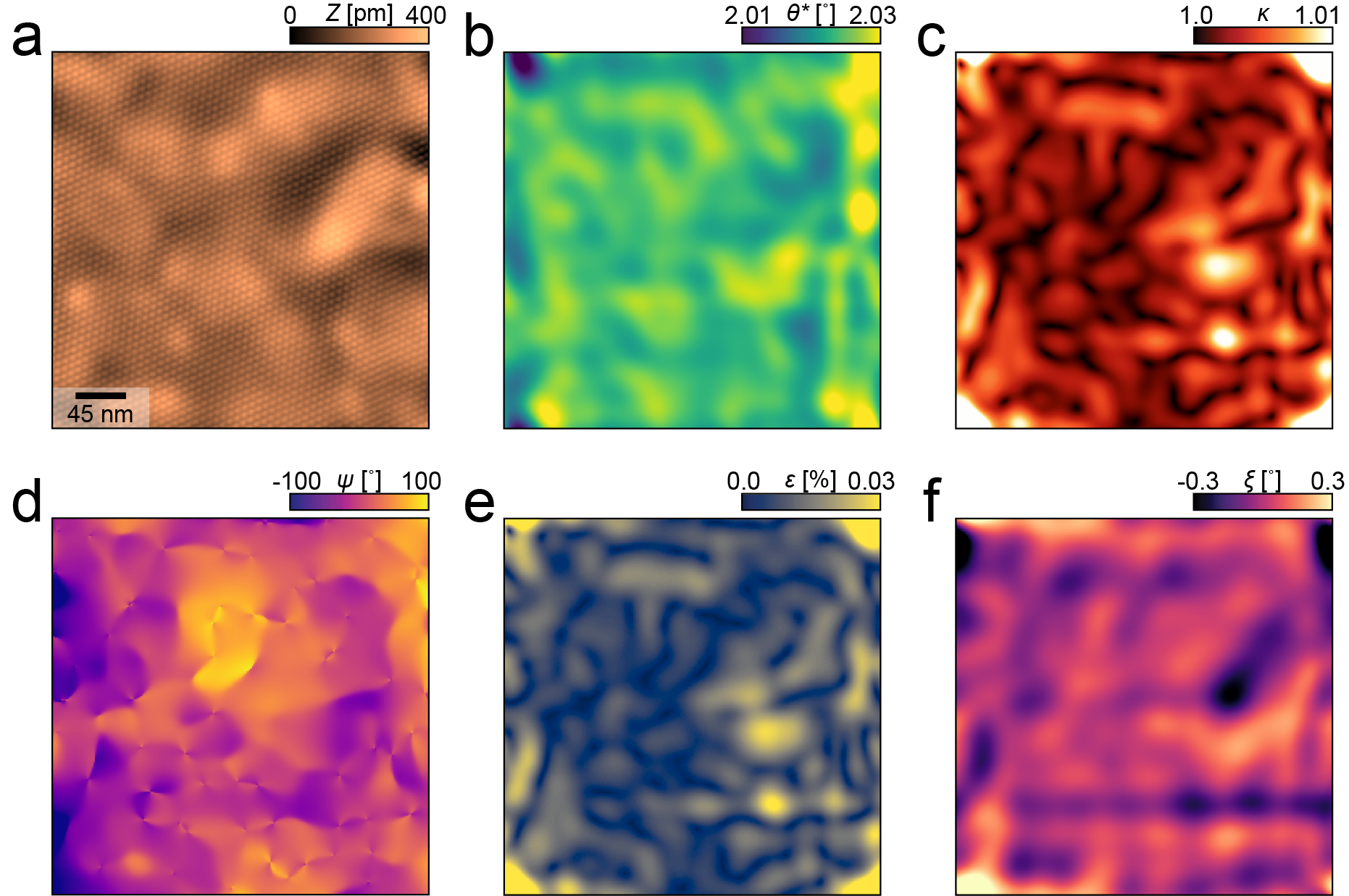}
\caption{a) Lawler--Fujita corrected STM topography of figure \ref{S2}a (and main text figure 4a).  b) Extracted effective twist angle map of a. c) Extracted residual local anisotropy map of a. d) Local anisotropy angle of c. e) Heterostrain map of a. f) Local moiré rotation of a. This angle corresponds to the angle in the $W$ matrix (equation \ref{eq:decomposition}).}
\label{S3}
\end{figure}

\noindent As an additional consistency check, we used the Lawler--Fujita algorithm  to reconstruct the undistorted image~\cite{n32}, and then applied the algorithm on the undistorted image in order to extract the residual displacement field and compare it to the previously extracted displacement field. Here, a perfectly performing and consistent algorithm would extract a zero residual displacement field. Therefore, this gives an indication of the error of the quantities extracted by the algorithm. Since we decompose the displacement field, for an almost zero displacement field, we expect the effective twist angle map to become more centered around the average twist angle (in this case, 2.02$^\circ$). Furthermore, we expected that most of the anisotropy is gone i.e., $\kappa \rightarrow 1$ and $\epsilon \rightarrow 0$.\\
We check this using the topography presented in the main text (figure 4) and in figure \ref{S2}a, and show the results in figure \ref{S3}. Aside from edge effects in the corner, both the residual anisotropy and the residual variations in the twist angle are more than an order of magnitude smaller than the originally obtained values, indicating self-consistency of the algorithm.
\FloatBarrier
\section{Validity check with more data}
\begin{figure*}[!hb]
    \centering
    \includegraphics[width=\linewidth]{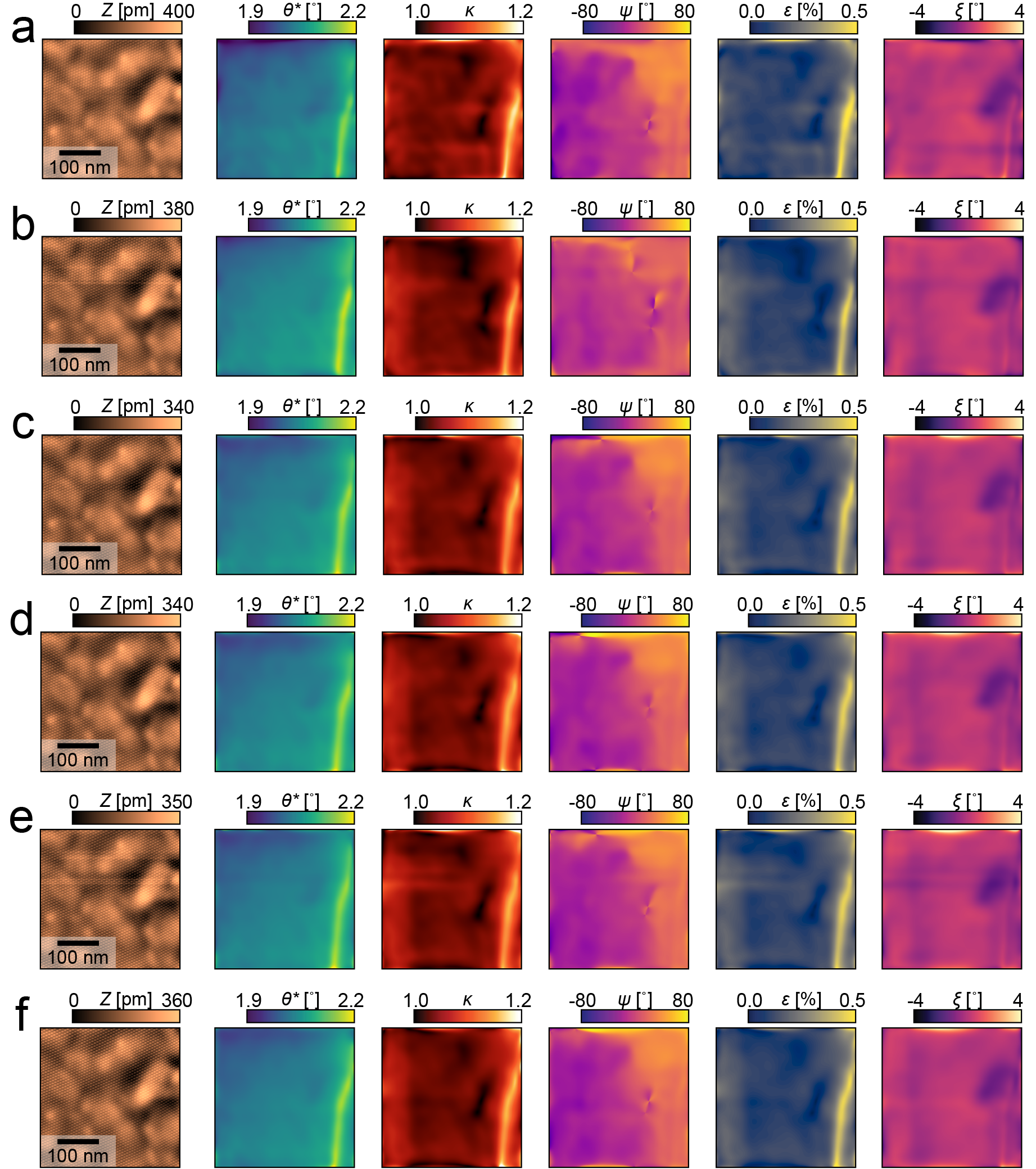}
    \caption{Spatial lock-in output for sequentially measured topographies in the same field of view. Figure f was measured at 65 nm/s, whereas a-e were measured with a scan speed of 54 nm/s. The setup condition was kept constant between measurements: V = 250 mV, I = 20 pA.}
    \label{S2}
\end{figure*}
In this work, we claim that the contribution of piezo drift to the output of our algorithm is negligible. 
To verify this, we apply it to multiple topographies, all sequentially measured on the same area.
All of them are measured with a scan speed of 54 nm/s, except the last one (figure \ref{S2}f), which is measured at 65 nm/s.
Because piezo drift changes with time and scan speed, comparing these datasets provides us with insight to which degree the algorithm output is affected by this effect.

The algorithm output for these measurements is displayed in figure \ref{S2}, where Figure \ref{S2}a corresponds to the data shown in the main text. For completeness, we also show $\xi(\vec r)$, the angle corresponding to the matrix $W$ (see section \ref{sec:deformations}).\\
Comparing these results from different scans, we observe that almost all deformations are reproduced, in particular the vertical line-like feature on the right and the two minima in $\kappa(\vec r)$. 
The only features not reproduced are horizontal `creases', corresponding to line-to-line scan artefacts.
Additionally, no significant difference is observed for figure \ref{S2}f with the deviating scan speed compared to the rest. 
From this, we conclude that most observed deformations are intrinsic to the sample.

\FloatBarrier

\section{Heterogeneity comparison with other devices and data overview}

\begin{figure}[!ht]
\centering
\includegraphics[width=\linewidth]{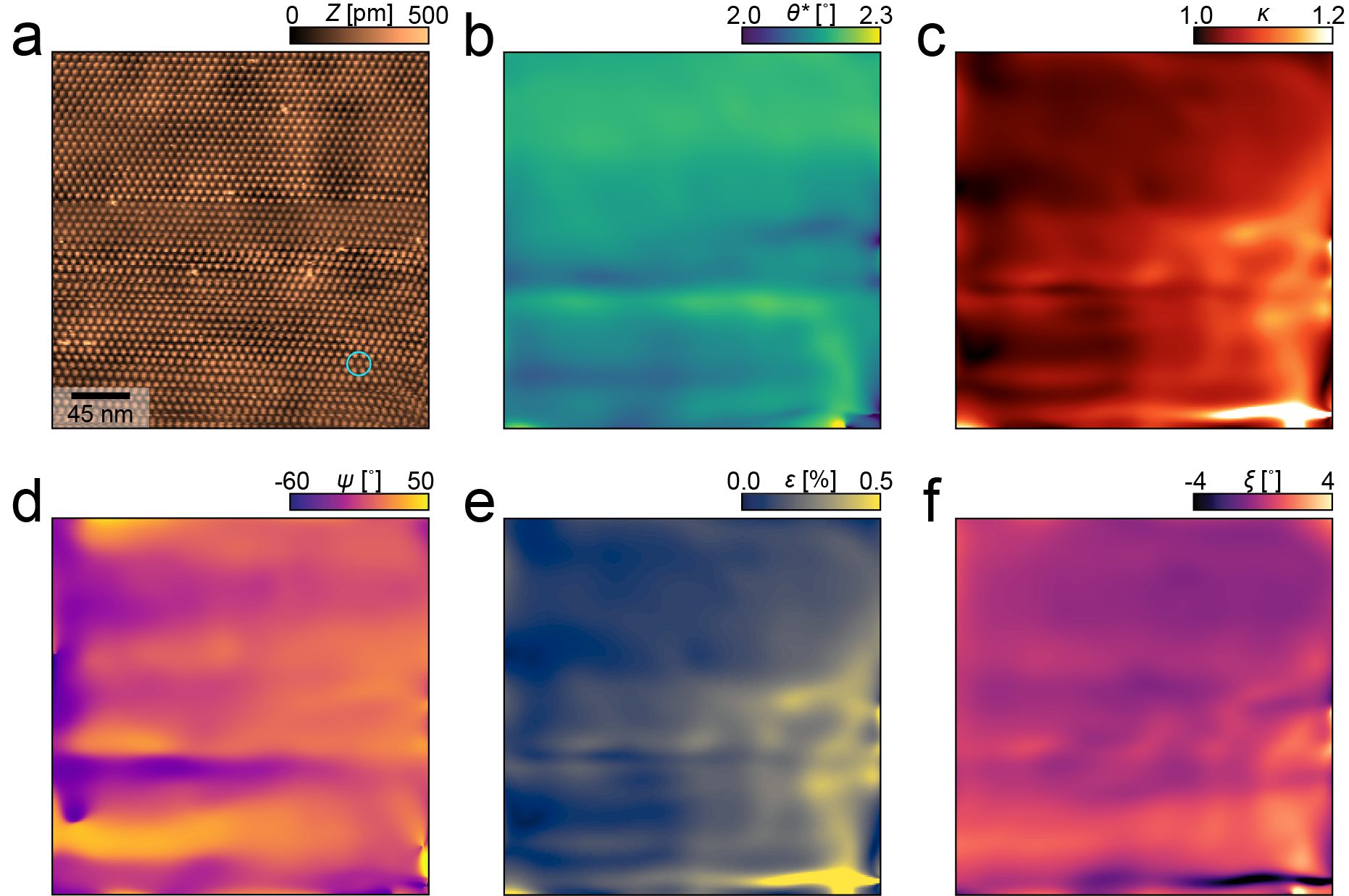}
\caption{a) STM topography of a TBG device with an average twist angle of 2.16$^\circ$ (set-up conditions: V = 170 mV, I = 20 pA). b) Extracted effective twist angle map of a. c) Extracted local anisotropy map of a. d) Local anisotropy angle of c. e) Heterostrain map of a. f) Local moiré rotation of a. This angle corresponds to the angle in the $W$ matrix (equation \ref{eq:decomposition}).}
\label{S4}
\end{figure}

\begin{figure}[!ht]
\centering
\includegraphics[width=\linewidth]{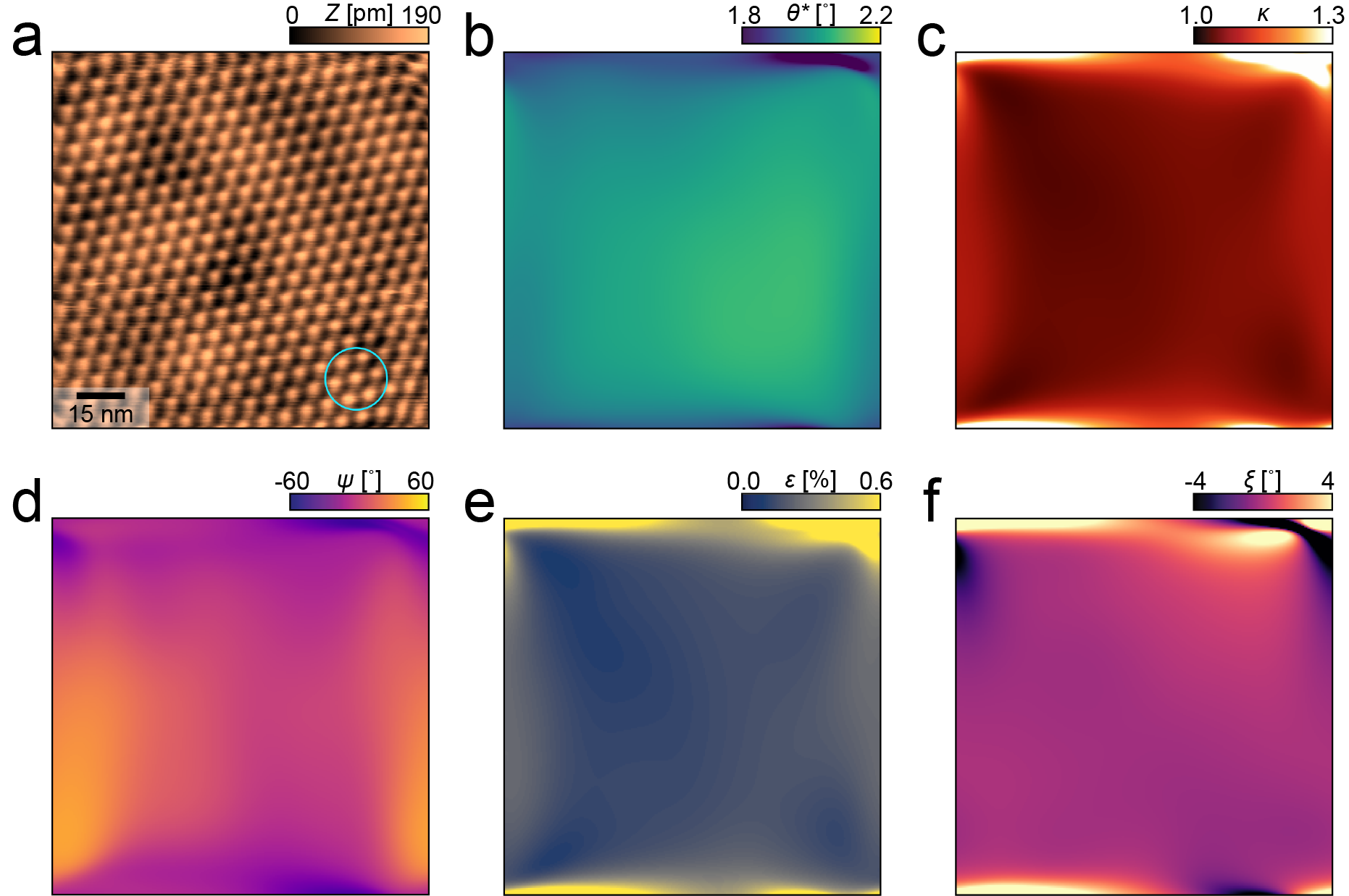}
\caption{a) STM topography of a TBG device with an average twist angle of 2.01$^\circ$ (set-up conditions: V = 350 mV, I = 100 pA).  b) Extracted effective twist angle map of a. c) Extracted local anisotropy map of a. d) Local anisotropy angle of c. e) Heterostrain map of a. f) Local moiré rotation of a. This angle corresponds to the angle in the $W$ matrix (equation \ref{eq:decomposition}).}
\label{S5}
\end{figure}

We measured 2 additional devices, with average twist angles of 2.16$^\circ$ and 2.01$^\circ$. The output of the spatial lock-in algorithm for these topographies is displayed in figure \ref{S4} and figure \ref{S5}. 
Calculating the standard deviation for the twist angle maps, we find 0.03$^\circ$ and 0.06$^\circ$ respectively, which is consistent with the result presented in the main text.\\

In total, topographies from 3 different devices are presented in this work. We give a short overview of the data measured per device, and the number of pixels for each measurement in table \ref{T1}
\begin{center}
\centering
\begin{table}[!ht]
\begin{tabular}{| c | c | c |}
\hline
Device & Figure & \# pixels \\
\hline
1 & 2a & $321\times321$ \\
2 & 2c, 4, 7, 8a, 11 & $246\times246$ \\
2 & 8b-f & $984\times984$ \\
2$^*$ & 10 & $256\times256$ \\
3 & 9 & $236\times236$ \\
\hline
\end{tabular}
\caption{Overview of the measured data per device.\\ $^*$: Measured after moving a few micrometer from the initial field of view.}
\label{T1}
\end{table}
\end{center}

\section{Homogeneity quantification}
\begin{figure}[!ht]
\centering
\includegraphics[width=\linewidth]{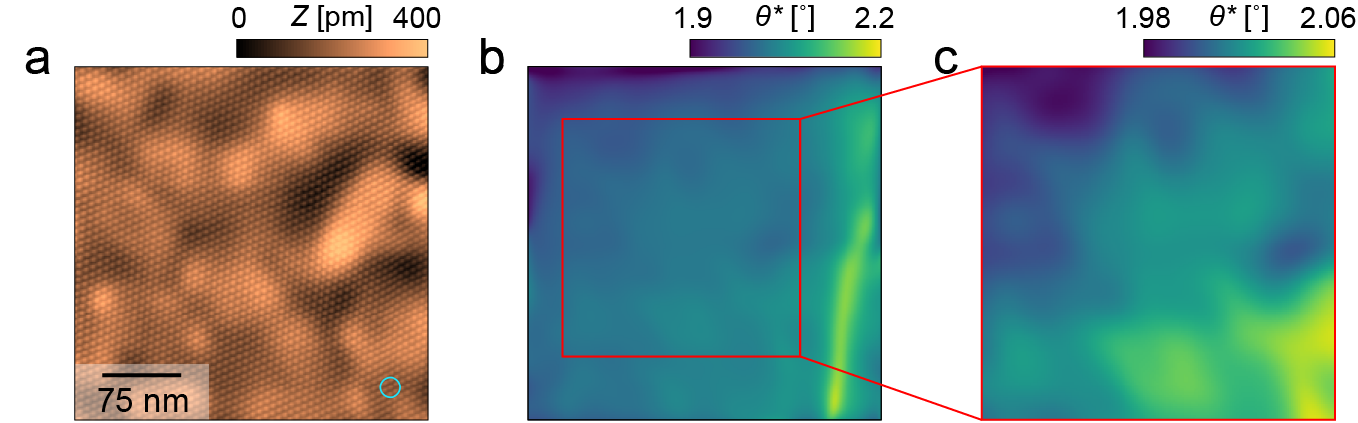}
\caption{a) STM topography of a device with an average twist angle of 2.02$^\circ$ (V = 250 mV, I = 20 pA, same data as main text figure 4a. b) Effective twist angle map extracted from a, by the algorithm discussed. The red square indicates the area over which the average twist angle and standard deviation are calculated. c) Effective twist angle map corresponding to the area marked by the red square in b.}
\label{S7}
\end{figure}

In the main text, the average, standard deviation and peak to peak spread are given for a cropped area of the twist angle map displayed in figure 4c. We find an average twist angle of 2.02$^\circ$ with a standard deviation of 0.02$^\circ$ and a peak to peak variation of 0.08$^\circ$. In figure \ref{S7}, we show the topography, along with the extracted twist angle map and the crop over which these values are calculated.\\

\begin{figure}[!ht]
\centering
\includegraphics[width=\linewidth]{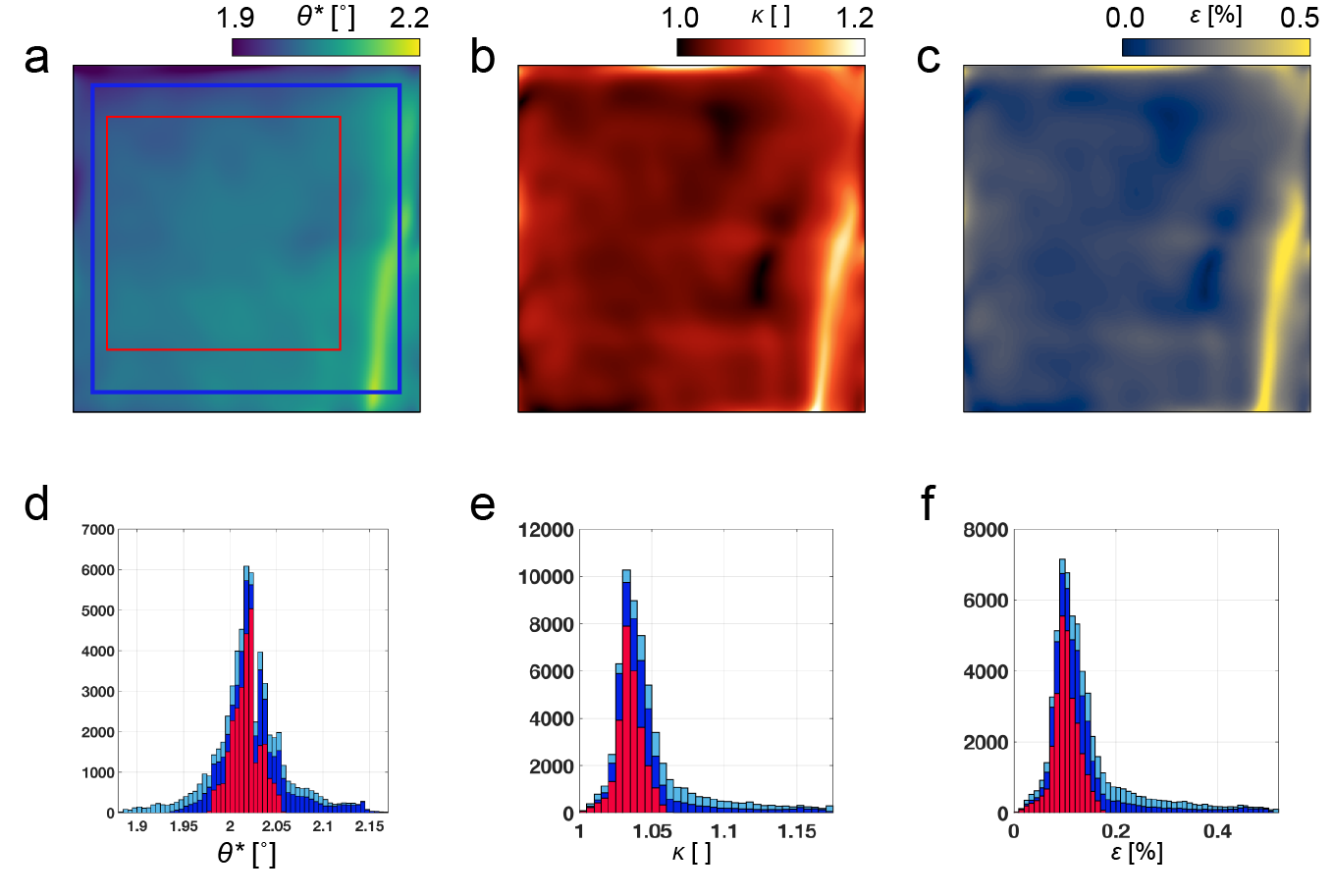}
\caption{a) Effective twist angle map extracted from the data shown in the main text, figure 4a. b) Local moir\'e anisotropy map $\kappa(\vec{r})$ extracted from the data shown in the main text, figure 4a. c) Heterostrain map extracted from the data shown in the main text, figure 4a. d-f) Histograms of the maps displayed above. The light blue histograms count the full data as shown, whereas the dark blue histograms exclude a border of two times the filter radius used in the lock-in procedure (pixels outside of the dark blue border in a). Finally, the red histograms count the data inside the area marked by the red square in a}
\label{sicrop}
\end{figure}

Another thing to consider here is some border effects that appear in our data. Because our method is based on lock-in techniques and the real space resolution is determined by the size of the filter we choose (see main text), we can expect an area around the border of our images to be affected by artifacts. We consider the size of this border to be about two times the radius of the filter used for the spatial lock-in procedure, motivated by the Gaussian profile of the filter: Two times the sigma of a Gaussian covers roughly 95\% of the weight of the window. In figure \ref{sicrop}, we show the extracted twist angle map, the local anisotropy and the heterostrain map of figure 4a in the main text, accompanied by histograms of each map both including and excluding the border.

\section{Artificial resolution limitation for comparison with SOT}
\begin{figure}[!ht]
\centering
\includegraphics[width=\linewidth]{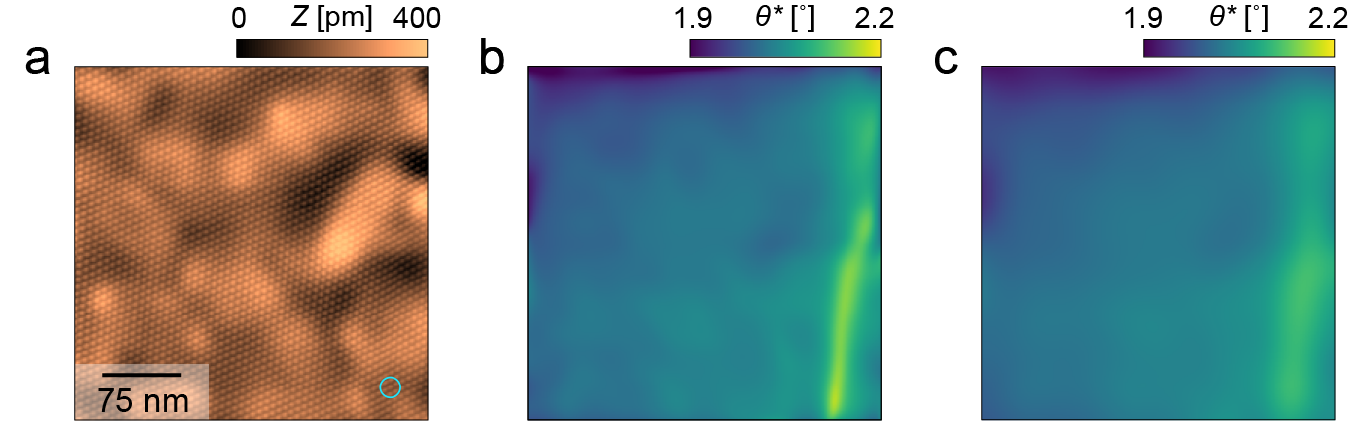}
\caption{a) STM topography of a device with an average twist angle of 2.02$^\circ$ (V = 250 mV, I = 20 pA, same data as main text figure 4a. b) Effective twist angle map extracted from a, by the algorithm discussed. c) Twist angle map of b after smearing with a Gaussian filter with a sigma of 15 nm.}
\label{S8}
\end{figure}

In order to compare our results with the result about twist angle homogeneity obtained on encapsulated device from SQUID-on-Tip (SOT) measurements \cite{n14}, we have to consider the resolution of SOT ($\sim$30 nm). To this end, we smear our obtained twist angle map (main text, figure 4c) with a Gaussian filter with a width of $\sigma = 15$ nm, the result of which is shown in figure \ref{S8}. Then, we obtain a peak to peak value of 0.20$^\circ$ and a standard deviation of 0.036$^\circ$.

\section{Error estimation of the heterostrain model}
\begin{figure}[!ht]
\centering
\includegraphics[width=\linewidth]{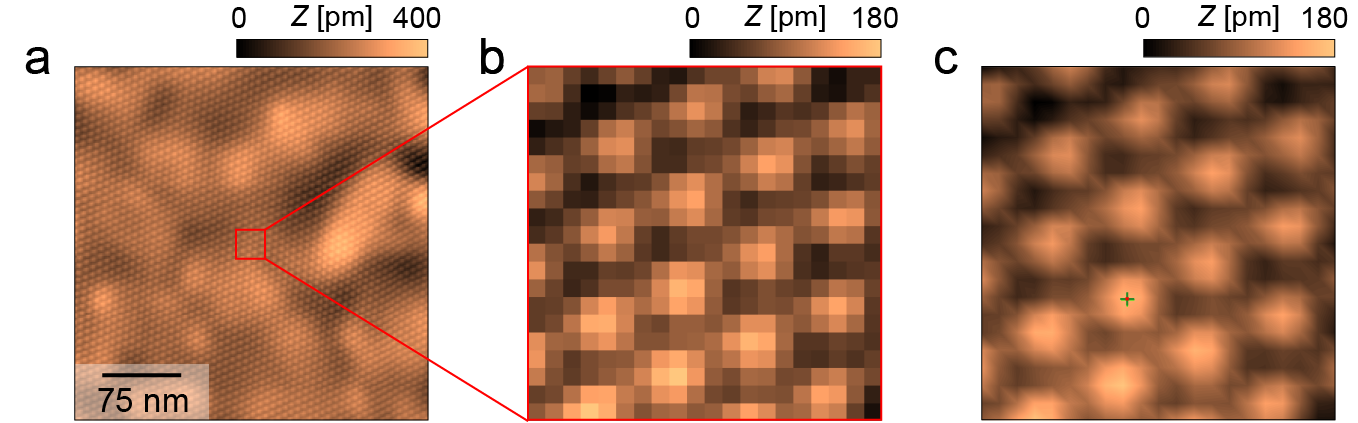}
\caption{a) STM topography of a device with an average twist angle of 2.02$^\circ$ (V = 250 mV, I = 20 pA, same data as main text figure 4a. b) Zoom in of a. c) Same figure as b, but plotted with interpolative shading. The red marker indicates the position of the Gaussian fitted to that moir\'e site and the green cross represents the extend 95\% confidence interval.}
\label{S9}
\end{figure}

In previous STM work, the twist angle of twisted bilayer graphene has been extracted with a heterostrain model \cite{n22}. This model relies heavily on accurately fitting a Gaussian to the moir\'e lattice sites in order to extract their position in space, and thereby, the relative distance between neighboring sites. Here, a Gaussian is fitted to a representative example of the moir\'e sites in our data and we calculate the 95\% confidence interval. The result is shown in figure \ref{S9}. We find a diameter of $\sim$1 nm for the 95\% confidence interval.

\section{Twist angle homogeneity overview}
In this section, we provide the extracted twist angle homogeneity of each device presented in this work (table \ref{T2}).


\begin{center}
\centering
\begin{table}[!ht]
\begin{tabularx}{\columnwidth}{| X | c | c | c |}
\hline
Description & Average  & peak-to-peak & standard  \\
& twist angle  &  variation & deviation  \\
\hline
\hline
Full field of view, figure 12 & 2.02$^\circ$ & 0.29$^\circ$ & 0.039$^\circ$ \\
Blue area, figure 12 & 2.02$^\circ$ & 0.23$^\circ$ & 0.033$^\circ$ \\
Red area, figure 12 & 2.02$^\circ$ & 0.08$^\circ$ & 0.015$^\circ$ \\
Full field of view,\newline figure 12,\newline after correcting PSF & 2.02$^\circ$ & 0.20$^\circ$ & 0.036$^\circ$ \\
\hline
Full field of view,\newline figure 9\newline (Blue area equivalent) & 2.16$^\circ$ & 0.22$^\circ$ & 0.029$^\circ$\\
\hline
Full field of view,\newline figure 10\newline (Blue area equivalent) & 2.03$^\circ$ & 0.14$^\circ$ & 0.029$^\circ$\\
\hline
\end{tabularx}
\caption{Overview of the twist angle homogeneity extracted per device.}
\label{T2}
\end{table}
\end{center}

\section{Data processing}
Regarding data pre-processing and post-processing, we made the following manipulations:
\begin{itemize}
\item Topographies are obtained from the measured data by subtracting a polynomial background up to $8^\text{th}$ order. It was verified that this did not significantly influence the extracted displacement fields.
\item The topography in figure 2a was additionally line subtracted.
\item FFT’s are calculated from the periodic part of the data, after applying the periodic + smooth decomposition algorithm~\cite{Moisan2010}.
\item The FFT in figure 4b uses interpolative shading.
\item The FFT in figure 2b is furthermore smeared with a gaussian filter (with a width of $\sigma = 0.5$ pixels).
\end{itemize}


\begin{thebibliography}{43}%
\makeatletter
\providecommand \@ifxundefined [1]{%
 \@ifx{#1\undefined}
}%
\providecommand \@ifnum [1]{%
 \ifnum #1\expandafter \@firstoftwo
 \else \expandafter \@secondoftwo
 \fi
}%
\providecommand \@ifx [1]{%
 \ifx #1\expandafter \@firstoftwo
 \else \expandafter \@secondoftwo
 \fi
}%
\providecommand \natexlab [1]{#1}%
\providecommand \enquote  [1]{``#1''}%
\providecommand \bibnamefont  [1]{#1}%
\providecommand \bibfnamefont [1]{#1}%
\providecommand \citenamefont [1]{#1}%
\providecommand \href@noop [0]{\@secondoftwo}%
\providecommand \href [0]{\begingroup \@sanitize@url \@href}%
\providecommand \@href[1]{\@@startlink{#1}\@@href}%
\providecommand \@@href[1]{\endgroup#1\@@endlink}%
\providecommand \@sanitize@url [0]{\catcode `\\12\catcode `\$12\catcode
  `\&12\catcode `\#12\catcode `\^12\catcode `\_12\catcode `\%12\relax}%
\providecommand \@@startlink[1]{}%
\providecommand \@@endlink[0]{}%
\providecommand \url  [0]{\begingroup\@sanitize@url \@url }%
\providecommand \@url [1]{\endgroup\@href {#1}{\urlprefix }}%
\providecommand \urlprefix  [0]{URL }%
\providecommand \Eprint [0]{\href }%
\providecommand \doibase [0]{https://doi.org/}%
\providecommand \selectlanguage [0]{\@gobble}%
\providecommand \bibinfo  [0]{\@secondoftwo}%
\providecommand \bibfield  [0]{\@secondoftwo}%
\providecommand \translation [1]{[#1]}%
\providecommand \BibitemOpen [0]{}%
\providecommand \bibitemStop [0]{}%
\providecommand \bibitemNoStop [0]{.\EOS\space}%
\providecommand \EOS [0]{\spacefactor3000\relax}%
\providecommand \BibitemShut  [1]{\csname bibitem#1\endcsname}%
\let\auto@bib@innerbib\@empty
\bibitem [{\citenamefont {Bistritzer}\ and\ \citenamefont
  {MacDonald}(2011)}]{n1}%
  \BibitemOpen
  \bibfield  {author} {\bibinfo {author} {\bibfnamefont {R.}~\bibnamefont
  {Bistritzer}}\ and\ \bibinfo {author} {\bibfnamefont {A.~H.}\ \bibnamefont
  {MacDonald}},\ }\bibfield  {title} {\bibinfo {title} {Moir{\'e} bands in
  twisted double-layer graphene},\ }\href
  {https://doi.org/10.1073/pnas.1108174108} {\bibfield  {journal} {\bibinfo
  {journal} {Proceedings of the National Academy of Sciences}\ }\textbf
  {\bibinfo {volume} {108}},\ \bibinfo {pages} {12233} (\bibinfo {year}
  {2011})}\BibitemShut {NoStop}%
\bibitem [{\citenamefont {Allan}\ \emph {et~al.}(2017)\citenamefont {Allan},
  \citenamefont {Fischer}, \citenamefont {Ostojic},\ and\ \citenamefont
  {Andringa}}]{n2}%
  \BibitemOpen
  \bibfield  {author} {\bibinfo {author} {\bibfnamefont {M.~P.}\ \bibnamefont
  {Allan}}, \bibinfo {author} {\bibfnamefont {M.~H.}\ \bibnamefont {Fischer}},
  \bibinfo {author} {\bibfnamefont {O.}~\bibnamefont {Ostojic}},\ and\ \bibinfo
  {author} {\bibfnamefont {A.}~\bibnamefont {Andringa}},\ }\bibfield  {title}
  {\bibinfo {title} {{Creating better superconductors by periodic
  nanopatterning}},\ }\href {https://doi.org/10.21468/SciPostPhys.3.2.010}
  {\bibfield  {journal} {\bibinfo  {journal} {SciPost Phys.}\ }\textbf
  {\bibinfo {volume} {3}},\ \bibinfo {pages} {010} (\bibinfo {year}
  {2017})}\BibitemShut {NoStop}%
\bibitem [{\citenamefont {Cao}\ \emph {et~al.}(2018{\natexlab{a}})\citenamefont
  {Cao}, \citenamefont {Fatemi}, \citenamefont {Demir}, \citenamefont {Fang},
  \citenamefont {Tomarken}, \citenamefont {Luo}, \citenamefont
  {Sanchez-Yamagishi}, \citenamefont {Watanabe}, \citenamefont {Taniguchi},
  \citenamefont {Kaxiras}, \citenamefont {Ashoori},\ and\ \citenamefont
  {Jarillo-Herrero}}]{n3}%
  \BibitemOpen
  \bibfield  {author} {\bibinfo {author} {\bibfnamefont {Y.}~\bibnamefont
  {Cao}}, \bibinfo {author} {\bibfnamefont {V.}~\bibnamefont {Fatemi}},
  \bibinfo {author} {\bibfnamefont {A.}~\bibnamefont {Demir}}, \bibinfo
  {author} {\bibfnamefont {S.}~\bibnamefont {Fang}}, \bibinfo {author}
  {\bibfnamefont {S.~L.}\ \bibnamefont {Tomarken}}, \bibinfo {author}
  {\bibfnamefont {J.~Y.}\ \bibnamefont {Luo}}, \bibinfo {author} {\bibfnamefont
  {J.~D.}\ \bibnamefont {Sanchez-Yamagishi}}, \bibinfo {author} {\bibfnamefont
  {K.}~\bibnamefont {Watanabe}}, \bibinfo {author} {\bibfnamefont
  {T.}~\bibnamefont {Taniguchi}}, \bibinfo {author} {\bibfnamefont
  {E.}~\bibnamefont {Kaxiras}}, \bibinfo {author} {\bibfnamefont {R.~C.}\
  \bibnamefont {Ashoori}},\ and\ \bibinfo {author} {\bibfnamefont
  {P.}~\bibnamefont {Jarillo-Herrero}},\ }\bibfield  {title} {\bibinfo {title}
  {Correlated insulator behaviour at half-filling in magic-angle graphene
  superlattices},\ }\href {https://doi.org/10.1038/nature26154} {\bibfield
  {journal} {\bibinfo  {journal} {Nature}\ }\textbf {\bibinfo {volume} {556}},\
  \bibinfo {pages} {80} (\bibinfo {year} {2018}{\natexlab{a}})}\BibitemShut
  {NoStop}%
\bibitem [{\citenamefont {Cao}\ \emph {et~al.}(2018{\natexlab{b}})\citenamefont
  {Cao}, \citenamefont {Fatemi}, \citenamefont {Fang}, \citenamefont
  {Watanabe}, \citenamefont {Taniguchi}, \citenamefont {Kaxiras},\ and\
  \citenamefont {Jarillo-Herrero}}]{n4}%
  \BibitemOpen
  \bibfield  {author} {\bibinfo {author} {\bibfnamefont {Y.}~\bibnamefont
  {Cao}}, \bibinfo {author} {\bibfnamefont {V.}~\bibnamefont {Fatemi}},
  \bibinfo {author} {\bibfnamefont {S.}~\bibnamefont {Fang}}, \bibinfo {author}
  {\bibfnamefont {K.}~\bibnamefont {Watanabe}}, \bibinfo {author}
  {\bibfnamefont {T.}~\bibnamefont {Taniguchi}}, \bibinfo {author}
  {\bibfnamefont {E.}~\bibnamefont {Kaxiras}},\ and\ \bibinfo {author}
  {\bibfnamefont {P.}~\bibnamefont {Jarillo-Herrero}},\ }\bibfield  {title}
  {\bibinfo {title} {Unconventional superconductivity in magic-angle graphene
  superlattices},\ }\href {https://doi.org/10.1038/nature26160} {\bibfield
  {journal} {\bibinfo  {journal} {Nature}\ }\textbf {\bibinfo {volume} {556}},\
  \bibinfo {pages} {43} (\bibinfo {year} {2018}{\natexlab{b}})}\BibitemShut
  {NoStop}%
\bibitem [{\citenamefont {Lu}\ \emph {et~al.}(2019)\citenamefont {Lu},
  \citenamefont {Stepanov}, \citenamefont {Yang}, \citenamefont {Xie},
  \citenamefont {Aamir}, \citenamefont {Das}, \citenamefont {Urgell},
  \citenamefont {Watanabe}, \citenamefont {Taniguchi}, \citenamefont {Zhang},
  \citenamefont {Bachtold}, \citenamefont {MacDonald},\ and\ \citenamefont
  {Efetov}}]{n5}%
  \BibitemOpen
  \bibfield  {author} {\bibinfo {author} {\bibfnamefont {X.}~\bibnamefont
  {Lu}}, \bibinfo {author} {\bibfnamefont {P.}~\bibnamefont {Stepanov}},
  \bibinfo {author} {\bibfnamefont {W.}~\bibnamefont {Yang}}, \bibinfo {author}
  {\bibfnamefont {M.}~\bibnamefont {Xie}}, \bibinfo {author} {\bibfnamefont
  {M.~A.}\ \bibnamefont {Aamir}}, \bibinfo {author} {\bibfnamefont
  {I.}~\bibnamefont {Das}}, \bibinfo {author} {\bibfnamefont {C.}~\bibnamefont
  {Urgell}}, \bibinfo {author} {\bibfnamefont {K.}~\bibnamefont {Watanabe}},
  \bibinfo {author} {\bibfnamefont {T.}~\bibnamefont {Taniguchi}}, \bibinfo
  {author} {\bibfnamefont {G.}~\bibnamefont {Zhang}}, \bibinfo {author}
  {\bibfnamefont {A.}~\bibnamefont {Bachtold}}, \bibinfo {author}
  {\bibfnamefont {A.~H.}\ \bibnamefont {MacDonald}},\ and\ \bibinfo {author}
  {\bibfnamefont {D.~K.}\ \bibnamefont {Efetov}},\ }\bibfield  {title}
  {\bibinfo {title} {Superconductors, orbital magnets and correlated states in
  magic-angle bilayer graphene},\ }\href
  {https://doi.org/10.1038/s41586-019-1695-0} {\bibfield  {journal} {\bibinfo
  {journal} {Nature}\ }\textbf {\bibinfo {volume} {574}},\ \bibinfo {pages}
  {653} (\bibinfo {year} {2019})}\BibitemShut {NoStop}%
\bibitem [{\citenamefont {Polshyn}\ \emph {et~al.}(2019)\citenamefont
  {Polshyn}, \citenamefont {Yankowitz}, \citenamefont {Chen}, \citenamefont
  {Zhang}, \citenamefont {Watanabe}, \citenamefont {Taniguchi}, \citenamefont
  {Dean},\ and\ \citenamefont {Young}}]{n6}%
  \BibitemOpen
  \bibfield  {author} {\bibinfo {author} {\bibfnamefont {H.}~\bibnamefont
  {Polshyn}}, \bibinfo {author} {\bibfnamefont {M.}~\bibnamefont {Yankowitz}},
  \bibinfo {author} {\bibfnamefont {S.}~\bibnamefont {Chen}}, \bibinfo {author}
  {\bibfnamefont {Y.}~\bibnamefont {Zhang}}, \bibinfo {author} {\bibfnamefont
  {K.}~\bibnamefont {Watanabe}}, \bibinfo {author} {\bibfnamefont
  {T.}~\bibnamefont {Taniguchi}}, \bibinfo {author} {\bibfnamefont {C.~R.}\
  \bibnamefont {Dean}},\ and\ \bibinfo {author} {\bibfnamefont {A.~F.}\
  \bibnamefont {Young}},\ }\bibfield  {title} {\bibinfo {title} {Large
  linear-in-temperature resistivity in twisted bilayer graphene},\ }\href
  {https://doi.org/10.1038/s41567-019-0596-3} {\bibfield  {journal} {\bibinfo
  {journal} {Nature Physics}\ }\textbf {\bibinfo {volume} {15}},\ \bibinfo
  {pages} {1011} (\bibinfo {year} {2019})}\BibitemShut {NoStop}%
\bibitem [{\citenamefont {Sharpe}\ \emph {et~al.}(2019)\citenamefont {Sharpe},
  \citenamefont {Fox}, \citenamefont {Barnard}, \citenamefont {Finney},
  \citenamefont {Watanabe}, \citenamefont {Taniguchi}, \citenamefont
  {Kastner},\ and\ \citenamefont {Goldhaber-Gordon}}]{n7}%
  \BibitemOpen
  \bibfield  {author} {\bibinfo {author} {\bibfnamefont {A.~L.}\ \bibnamefont
  {Sharpe}}, \bibinfo {author} {\bibfnamefont {E.~J.}\ \bibnamefont {Fox}},
  \bibinfo {author} {\bibfnamefont {A.~W.}\ \bibnamefont {Barnard}}, \bibinfo
  {author} {\bibfnamefont {J.}~\bibnamefont {Finney}}, \bibinfo {author}
  {\bibfnamefont {K.}~\bibnamefont {Watanabe}}, \bibinfo {author}
  {\bibfnamefont {T.}~\bibnamefont {Taniguchi}}, \bibinfo {author}
  {\bibfnamefont {M.~A.}\ \bibnamefont {Kastner}},\ and\ \bibinfo {author}
  {\bibfnamefont {D.}~\bibnamefont {Goldhaber-Gordon}},\ }\bibfield  {title}
  {\bibinfo {title} {Emergent ferromagnetism near three-quarters filling in
  twisted bilayer graphene},\ }\href {https://doi.org/10.1126/science.aaw3780}
  {\bibfield  {journal} {\bibinfo  {journal} {Science}\ }\textbf {\bibinfo
  {volume} {365}},\ \bibinfo {pages} {605} (\bibinfo {year}
  {2019})}\BibitemShut {NoStop}%
\bibitem [{\citenamefont {Yankowitz}\ \emph {et~al.}(2019)\citenamefont
  {Yankowitz}, \citenamefont {Chen}, \citenamefont {Polshyn}, \citenamefont
  {Zhang}, \citenamefont {Watanabe}, \citenamefont {Taniguchi}, \citenamefont
  {Graf}, \citenamefont {Young},\ and\ \citenamefont {Dean}}]{n8}%
  \BibitemOpen
  \bibfield  {author} {\bibinfo {author} {\bibfnamefont {M.}~\bibnamefont
  {Yankowitz}}, \bibinfo {author} {\bibfnamefont {S.}~\bibnamefont {Chen}},
  \bibinfo {author} {\bibfnamefont {H.}~\bibnamefont {Polshyn}}, \bibinfo
  {author} {\bibfnamefont {Y.}~\bibnamefont {Zhang}}, \bibinfo {author}
  {\bibfnamefont {K.}~\bibnamefont {Watanabe}}, \bibinfo {author}
  {\bibfnamefont {T.}~\bibnamefont {Taniguchi}}, \bibinfo {author}
  {\bibfnamefont {D.}~\bibnamefont {Graf}}, \bibinfo {author} {\bibfnamefont
  {A.~F.}\ \bibnamefont {Young}},\ and\ \bibinfo {author} {\bibfnamefont
  {C.~R.}\ \bibnamefont {Dean}},\ }\bibfield  {title} {\bibinfo {title} {Tuning
  superconductivity in twisted bilayer graphene},\ }\href
  {https://doi.org/10.1126/science.aav1910} {\bibfield  {journal} {\bibinfo
  {journal} {Science}\ }\textbf {\bibinfo {volume} {363}},\ \bibinfo {pages}
  {1059} (\bibinfo {year} {2019})}\BibitemShut {NoStop}%
\bibitem [{\citenamefont {Stepanov}\ \emph {et~al.}(2020)\citenamefont
  {Stepanov}, \citenamefont {Das}, \citenamefont {Lu}, \citenamefont
  {Fahimniya}, \citenamefont {Watanabe}, \citenamefont {Taniguchi},
  \citenamefont {Koppens}, \citenamefont {Lischner}, \citenamefont {Levitov},\
  and\ \citenamefont {Efetov}}]{n9}%
  \BibitemOpen
  \bibfield  {author} {\bibinfo {author} {\bibfnamefont {P.}~\bibnamefont
  {Stepanov}}, \bibinfo {author} {\bibfnamefont {I.}~\bibnamefont {Das}},
  \bibinfo {author} {\bibfnamefont {X.}~\bibnamefont {Lu}}, \bibinfo {author}
  {\bibfnamefont {A.}~\bibnamefont {Fahimniya}}, \bibinfo {author}
  {\bibfnamefont {K.}~\bibnamefont {Watanabe}}, \bibinfo {author}
  {\bibfnamefont {T.}~\bibnamefont {Taniguchi}}, \bibinfo {author}
  {\bibfnamefont {F.~H.~L.}\ \bibnamefont {Koppens}}, \bibinfo {author}
  {\bibfnamefont {J.}~\bibnamefont {Lischner}}, \bibinfo {author}
  {\bibfnamefont {L.}~\bibnamefont {Levitov}},\ and\ \bibinfo {author}
  {\bibfnamefont {D.~K.}\ \bibnamefont {Efetov}},\ }\bibfield  {title}
  {\bibinfo {title} {Untying the insulating and superconducting orders in
  magic-angle graphene},\ }\href {https://doi.org/10.1038/s41586-020-2459-6}
  {\bibfield  {journal} {\bibinfo  {journal} {Nature}\ }\textbf {\bibinfo
  {volume} {583}},\ \bibinfo {pages} {375} (\bibinfo {year}
  {2020})}\BibitemShut {NoStop}%
\bibitem [{\citenamefont {Codecido}\ \emph {et~al.}(2019)\citenamefont
  {Codecido}, \citenamefont {Wang}, \citenamefont {Koester}, \citenamefont
  {Che}, \citenamefont {Tian}, \citenamefont {Lv}, \citenamefont {Tran},
  \citenamefont {Watanabe}, \citenamefont {Taniguchi}, \citenamefont {Zhang},
  \citenamefont {Bockrath},\ and\ \citenamefont {Lau}}]{n10}%
  \BibitemOpen
  \bibfield  {author} {\bibinfo {author} {\bibfnamefont {E.}~\bibnamefont
  {Codecido}}, \bibinfo {author} {\bibfnamefont {Q.}~\bibnamefont {Wang}},
  \bibinfo {author} {\bibfnamefont {R.}~\bibnamefont {Koester}}, \bibinfo
  {author} {\bibfnamefont {S.}~\bibnamefont {Che}}, \bibinfo {author}
  {\bibfnamefont {H.}~\bibnamefont {Tian}}, \bibinfo {author} {\bibfnamefont
  {R.}~\bibnamefont {Lv}}, \bibinfo {author} {\bibfnamefont {S.}~\bibnamefont
  {Tran}}, \bibinfo {author} {\bibfnamefont {K.}~\bibnamefont {Watanabe}},
  \bibinfo {author} {\bibfnamefont {T.}~\bibnamefont {Taniguchi}}, \bibinfo
  {author} {\bibfnamefont {F.}~\bibnamefont {Zhang}}, \bibinfo {author}
  {\bibfnamefont {M.}~\bibnamefont {Bockrath}},\ and\ \bibinfo {author}
  {\bibfnamefont {C.~N.}\ \bibnamefont {Lau}},\ }\bibfield  {title} {\bibinfo
  {title} {Correlated insulating and superconducting states in twisted bilayer
  graphene below the magic angle},\ }\bibfield  {journal} {\bibinfo  {journal}
  {Science Advances}\ }\textbf {\bibinfo {volume} {5}},\ \href
  {https://doi.org/10.1126/sciadv.aaw9770} {10.1126/sciadv.aaw9770} (\bibinfo
  {year} {2019})\BibitemShut {NoStop}%
\bibitem [{\citenamefont {Saito}\ \emph {et~al.}(2020)\citenamefont {Saito},
  \citenamefont {Ge}, \citenamefont {Watanabe}, \citenamefont {Taniguchi},\
  and\ \citenamefont {Young}}]{n11}%
  \BibitemOpen
  \bibfield  {author} {\bibinfo {author} {\bibfnamefont {Y.}~\bibnamefont
  {Saito}}, \bibinfo {author} {\bibfnamefont {J.}~\bibnamefont {Ge}}, \bibinfo
  {author} {\bibfnamefont {K.}~\bibnamefont {Watanabe}}, \bibinfo {author}
  {\bibfnamefont {T.}~\bibnamefont {Taniguchi}},\ and\ \bibinfo {author}
  {\bibfnamefont {A.~F.}\ \bibnamefont {Young}},\ }\bibfield  {title} {\bibinfo
  {title} {Independent superconductors and correlated insulators in twisted
  bilayer graphene},\ }\href {https://doi.org/10.1038/s41567-020-0928-3}
  {\bibfield  {journal} {\bibinfo  {journal} {Nature Physics}\ }\textbf
  {\bibinfo {volume} {16}},\ \bibinfo {pages} {926} (\bibinfo {year}
  {2020})}\BibitemShut {NoStop}%
\bibitem [{\citenamefont {Cao}\ \emph {et~al.}(2020)\citenamefont {Cao},
  \citenamefont {Chowdhury}, \citenamefont {Rodan-Legrain}, \citenamefont
  {Rubies-Bigorda}, \citenamefont {Watanabe}, \citenamefont {Taniguchi},
  \citenamefont {Senthil},\ and\ \citenamefont {Jarillo-Herrero}}]{n12}%
  \BibitemOpen
  \bibfield  {author} {\bibinfo {author} {\bibfnamefont {Y.}~\bibnamefont
  {Cao}}, \bibinfo {author} {\bibfnamefont {D.}~\bibnamefont {Chowdhury}},
  \bibinfo {author} {\bibfnamefont {D.}~\bibnamefont {Rodan-Legrain}}, \bibinfo
  {author} {\bibfnamefont {O.}~\bibnamefont {Rubies-Bigorda}}, \bibinfo
  {author} {\bibfnamefont {K.}~\bibnamefont {Watanabe}}, \bibinfo {author}
  {\bibfnamefont {T.}~\bibnamefont {Taniguchi}}, \bibinfo {author}
  {\bibfnamefont {T.}~\bibnamefont {Senthil}},\ and\ \bibinfo {author}
  {\bibfnamefont {P.}~\bibnamefont {Jarillo-Herrero}},\ }\bibfield  {title}
  {\bibinfo {title} {Strange metal in magic-angle graphene with near planckian
  dissipation},\ }\href {https://doi.org/10.1103/PhysRevLett.124.076801}
  {\bibfield  {journal} {\bibinfo  {journal} {Phys. Rev. Lett.}\ }\textbf
  {\bibinfo {volume} {124}},\ \bibinfo {pages} {076801} (\bibinfo {year}
  {2020})}\BibitemShut {NoStop}%
\bibitem [{\citenamefont {Balents}\ \emph {et~al.}(2020)\citenamefont
  {Balents}, \citenamefont {Dean}, \citenamefont {Efetov},\ and\ \citenamefont
  {Young}}]{n13}%
  \BibitemOpen
  \bibfield  {author} {\bibinfo {author} {\bibfnamefont {L.}~\bibnamefont
  {Balents}}, \bibinfo {author} {\bibfnamefont {C.~R.}\ \bibnamefont {Dean}},
  \bibinfo {author} {\bibfnamefont {D.~K.}\ \bibnamefont {Efetov}},\ and\
  \bibinfo {author} {\bibfnamefont {A.~F.}\ \bibnamefont {Young}},\ }\bibfield
  {title} {\bibinfo {title} {Superconductivity and strong correlations in
  moir{\'e}flat bands},\ }\href {https://doi.org/10.1038/s41567-020-0906-9}
  {\bibfield  {journal} {\bibinfo  {journal} {Nature Physics}\ }\textbf
  {\bibinfo {volume} {16}},\ \bibinfo {pages} {725} (\bibinfo {year}
  {2020})}\BibitemShut {NoStop}%
\bibitem [{\citenamefont {Uri}\ \emph {et~al.}(2020)\citenamefont {Uri},
  \citenamefont {Grover}, \citenamefont {Cao}, \citenamefont {Crosse},
  \citenamefont {Bagani}, \citenamefont {Rodan-Legrain}, \citenamefont
  {Myasoedov}, \citenamefont {Watanabe}, \citenamefont {Taniguchi},
  \citenamefont {Moon}, \citenamefont {Koshino}, \citenamefont
  {Jarillo-Herrero},\ and\ \citenamefont {Zeldov}}]{n14}%
  \BibitemOpen
  \bibfield  {author} {\bibinfo {author} {\bibfnamefont {A.}~\bibnamefont
  {Uri}}, \bibinfo {author} {\bibfnamefont {S.}~\bibnamefont {Grover}},
  \bibinfo {author} {\bibfnamefont {Y.}~\bibnamefont {Cao}}, \bibinfo {author}
  {\bibfnamefont {J.~A.}\ \bibnamefont {Crosse}}, \bibinfo {author}
  {\bibfnamefont {K.}~\bibnamefont {Bagani}}, \bibinfo {author} {\bibfnamefont
  {D.}~\bibnamefont {Rodan-Legrain}}, \bibinfo {author} {\bibfnamefont
  {Y.}~\bibnamefont {Myasoedov}}, \bibinfo {author} {\bibfnamefont
  {K.}~\bibnamefont {Watanabe}}, \bibinfo {author} {\bibfnamefont
  {T.}~\bibnamefont {Taniguchi}}, \bibinfo {author} {\bibfnamefont
  {P.}~\bibnamefont {Moon}}, \bibinfo {author} {\bibfnamefont {M.}~\bibnamefont
  {Koshino}}, \bibinfo {author} {\bibfnamefont {P.}~\bibnamefont
  {Jarillo-Herrero}},\ and\ \bibinfo {author} {\bibfnamefont {E.}~\bibnamefont
  {Zeldov}},\ }\bibfield  {title} {\bibinfo {title} {Mapping the twist-angle
  disorder and landau levels in magic-angle graphene},\ }\href
  {https://doi.org/10.1038/s41586-020-2255-3} {\bibfield  {journal} {\bibinfo
  {journal} {Nature}\ }\textbf {\bibinfo {volume} {581}},\ \bibinfo {pages}
  {47} (\bibinfo {year} {2020})}\BibitemShut {NoStop}%
\bibitem [{\citenamefont {Lisi}\ \emph {et~al.}(2020)\citenamefont {Lisi},
  \citenamefont {Lu}, \citenamefont {Benschop}, \citenamefont {de~Jong},
  \citenamefont {Stepanov}, \citenamefont {Duran}, \citenamefont {Margot},
  \citenamefont {Cucchi}, \citenamefont {Cappelli}, \citenamefont {Hunter},
  \citenamefont {Tamai}, \citenamefont {Kandyba}, \citenamefont {Giampietri},
  \citenamefont {Barinov}, \citenamefont {Jobst}, \citenamefont {Stalman},
  \citenamefont {Leeuwenhoek}, \citenamefont {Watanabe}, \citenamefont
  {Taniguchi}, \citenamefont {Rademaker}, \citenamefont {van~der Molen},
  \citenamefont {Allan}, \citenamefont {Efetov},\ and\ \citenamefont
  {Baumberger}}]{n15}%
  \BibitemOpen
  \bibfield  {author} {\bibinfo {author} {\bibfnamefont {S.}~\bibnamefont
  {Lisi}}, \bibinfo {author} {\bibfnamefont {X.}~\bibnamefont {Lu}}, \bibinfo
  {author} {\bibfnamefont {T.}~\bibnamefont {Benschop}}, \bibinfo {author}
  {\bibfnamefont {T.~A.}\ \bibnamefont {de~Jong}}, \bibinfo {author}
  {\bibfnamefont {P.}~\bibnamefont {Stepanov}}, \bibinfo {author}
  {\bibfnamefont {J.~R.}\ \bibnamefont {Duran}}, \bibinfo {author}
  {\bibfnamefont {F.}~\bibnamefont {Margot}}, \bibinfo {author} {\bibfnamefont
  {I.}~\bibnamefont {Cucchi}}, \bibinfo {author} {\bibfnamefont
  {E.}~\bibnamefont {Cappelli}}, \bibinfo {author} {\bibfnamefont
  {A.}~\bibnamefont {Hunter}}, \bibinfo {author} {\bibfnamefont
  {A.}~\bibnamefont {Tamai}}, \bibinfo {author} {\bibfnamefont
  {V.}~\bibnamefont {Kandyba}}, \bibinfo {author} {\bibfnamefont
  {A.}~\bibnamefont {Giampietri}}, \bibinfo {author} {\bibfnamefont
  {A.}~\bibnamefont {Barinov}}, \bibinfo {author} {\bibfnamefont
  {J.}~\bibnamefont {Jobst}}, \bibinfo {author} {\bibfnamefont
  {V.}~\bibnamefont {Stalman}}, \bibinfo {author} {\bibfnamefont
  {M.}~\bibnamefont {Leeuwenhoek}}, \bibinfo {author} {\bibfnamefont
  {K.}~\bibnamefont {Watanabe}}, \bibinfo {author} {\bibfnamefont
  {T.}~\bibnamefont {Taniguchi}}, \bibinfo {author} {\bibfnamefont
  {L.}~\bibnamefont {Rademaker}}, \bibinfo {author} {\bibfnamefont {S.~J.}\
  \bibnamefont {van~der Molen}}, \bibinfo {author} {\bibfnamefont
  {M.}~\bibnamefont {Allan}}, \bibinfo {author} {\bibfnamefont {D.~K.}\
  \bibnamefont {Efetov}},\ and\ \bibinfo {author} {\bibfnamefont
  {F.}~\bibnamefont {Baumberger}},\ }\href@noop {} {\bibinfo {title} {Direct
  evidence for flat bands in twisted bilayer graphene from nano-arpes}}
  (\bibinfo {year} {2020}),\ \Eprint {https://arxiv.org/abs/2002.02289}
  {arXiv:2002.02289 [cond-mat.str-el]} \BibitemShut {NoStop}%
\bibitem [{\citenamefont {Razado-Colambo}\ \emph {et~al.}(2016)\citenamefont
  {Razado-Colambo}, \citenamefont {Avila}, \citenamefont {Nys}, \citenamefont
  {Chen}, \citenamefont {Wallart}, \citenamefont {Asensio},\ and\ \citenamefont
  {Vignaud}}]{n16}%
  \BibitemOpen
  \bibfield  {author} {\bibinfo {author} {\bibfnamefont {I.}~\bibnamefont
  {Razado-Colambo}}, \bibinfo {author} {\bibfnamefont {J.}~\bibnamefont
  {Avila}}, \bibinfo {author} {\bibfnamefont {J.~P.}\ \bibnamefont {Nys}},
  \bibinfo {author} {\bibfnamefont {C.}~\bibnamefont {Chen}}, \bibinfo {author}
  {\bibfnamefont {X.}~\bibnamefont {Wallart}}, \bibinfo {author} {\bibfnamefont
  {M.~C.}\ \bibnamefont {Asensio}},\ and\ \bibinfo {author} {\bibfnamefont
  {D.}~\bibnamefont {Vignaud}},\ }\bibfield  {title} {\bibinfo {title}
  {Nanoarpes of twisted bilayer graphene on sic: absence of velocity
  renormalization for small angles},\ }\href
  {https://doi.org/10.1038/srep27261} {\bibfield  {journal} {\bibinfo
  {journal} {Scientific Reports}\ }\textbf {\bibinfo {volume} {6}},\ \bibinfo
  {pages} {27261} (\bibinfo {year} {2016})}\BibitemShut {NoStop}%
\bibitem [{\citenamefont {Jones}\ \emph {et~al.}(2020)\citenamefont {Jones},
  \citenamefont {Muzzio}, \citenamefont {Majchrzak}, \citenamefont {Pakdel},
  \citenamefont {Curcio}, \citenamefont {Volckaert}, \citenamefont {Biswas},
  \citenamefont {Gobbo}, \citenamefont {Singh}, \citenamefont {Robinson},
  \citenamefont {Watanabe}, \citenamefont {Taniguchi}, \citenamefont {Kim},
  \citenamefont {Cacho}, \citenamefont {Lanata}, \citenamefont {Miwa},
  \citenamefont {Hofmann}, \citenamefont {Katoch},\ and\ \citenamefont
  {Ulstrup}}]{n17}%
  \BibitemOpen
  \bibfield  {author} {\bibinfo {author} {\bibfnamefont {A.~J.~H.}\
  \bibnamefont {Jones}}, \bibinfo {author} {\bibfnamefont {R.}~\bibnamefont
  {Muzzio}}, \bibinfo {author} {\bibfnamefont {P.}~\bibnamefont {Majchrzak}},
  \bibinfo {author} {\bibfnamefont {S.}~\bibnamefont {Pakdel}}, \bibinfo
  {author} {\bibfnamefont {D.}~\bibnamefont {Curcio}}, \bibinfo {author}
  {\bibfnamefont {K.}~\bibnamefont {Volckaert}}, \bibinfo {author}
  {\bibfnamefont {D.}~\bibnamefont {Biswas}}, \bibinfo {author} {\bibfnamefont
  {J.}~\bibnamefont {Gobbo}}, \bibinfo {author} {\bibfnamefont
  {S.}~\bibnamefont {Singh}}, \bibinfo {author} {\bibfnamefont {J.~T.}\
  \bibnamefont {Robinson}}, \bibinfo {author} {\bibfnamefont {K.}~\bibnamefont
  {Watanabe}}, \bibinfo {author} {\bibfnamefont {T.}~\bibnamefont {Taniguchi}},
  \bibinfo {author} {\bibfnamefont {T.~K.}\ \bibnamefont {Kim}}, \bibinfo
  {author} {\bibfnamefont {C.}~\bibnamefont {Cacho}}, \bibinfo {author}
  {\bibfnamefont {N.}~\bibnamefont {Lanata}}, \bibinfo {author} {\bibfnamefont
  {J.~A.}\ \bibnamefont {Miwa}}, \bibinfo {author} {\bibfnamefont
  {P.}~\bibnamefont {Hofmann}}, \bibinfo {author} {\bibfnamefont
  {J.}~\bibnamefont {Katoch}},\ and\ \bibinfo {author} {\bibfnamefont
  {S.}~\bibnamefont {Ulstrup}},\ }\bibfield  {title} {\bibinfo {title}
  {Observation of electrically tunable van hove singularities in twisted
  bilayer graphene from nanoarpes},\ }\href
  {https://doi.org/10.1002/adma.202001656} {\bibfield  {journal} {\bibinfo
  {journal} {Advanced Materials}\ }\textbf {\bibinfo {volume} {32}},\ \bibinfo
  {pages} {2001656} (\bibinfo {year} {2020})}\BibitemShut {NoStop}%
\bibitem [{\citenamefont {McGilly}\ \emph {et~al.}(2020)\citenamefont
  {McGilly}, \citenamefont {Kerelsky}, \citenamefont {Finney}, \citenamefont
  {Shapovalov}, \citenamefont {Shih}, \citenamefont {Ghiotto}, \citenamefont
  {Zeng}, \citenamefont {Moore}, \citenamefont {Wu}, \citenamefont {Bai},
  \citenamefont {Watanabe}, \citenamefont {Taniguchi}, \citenamefont {Stengel},
  \citenamefont {Zhou}, \citenamefont {Hone}, \citenamefont {Zhu},
  \citenamefont {Basov}, \citenamefont {Dean}, \citenamefont {Dreyer},\ and\
  \citenamefont {Pasupathy}}]{n18}%
  \BibitemOpen
  \bibfield  {author} {\bibinfo {author} {\bibfnamefont {L.~J.}\ \bibnamefont
  {McGilly}}, \bibinfo {author} {\bibfnamefont {A.}~\bibnamefont {Kerelsky}},
  \bibinfo {author} {\bibfnamefont {N.~R.}\ \bibnamefont {Finney}}, \bibinfo
  {author} {\bibfnamefont {K.}~\bibnamefont {Shapovalov}}, \bibinfo {author}
  {\bibfnamefont {E.-M.}\ \bibnamefont {Shih}}, \bibinfo {author}
  {\bibfnamefont {A.}~\bibnamefont {Ghiotto}}, \bibinfo {author} {\bibfnamefont
  {Y.}~\bibnamefont {Zeng}}, \bibinfo {author} {\bibfnamefont {S.~L.}\
  \bibnamefont {Moore}}, \bibinfo {author} {\bibfnamefont {W.}~\bibnamefont
  {Wu}}, \bibinfo {author} {\bibfnamefont {Y.}~\bibnamefont {Bai}}, \bibinfo
  {author} {\bibfnamefont {K.}~\bibnamefont {Watanabe}}, \bibinfo {author}
  {\bibfnamefont {T.}~\bibnamefont {Taniguchi}}, \bibinfo {author}
  {\bibfnamefont {M.}~\bibnamefont {Stengel}}, \bibinfo {author} {\bibfnamefont
  {L.}~\bibnamefont {Zhou}}, \bibinfo {author} {\bibfnamefont {J.}~\bibnamefont
  {Hone}}, \bibinfo {author} {\bibfnamefont {X.}~\bibnamefont {Zhu}}, \bibinfo
  {author} {\bibfnamefont {D.~N.}\ \bibnamefont {Basov}}, \bibinfo {author}
  {\bibfnamefont {C.}~\bibnamefont {Dean}}, \bibinfo {author} {\bibfnamefont
  {C.~E.}\ \bibnamefont {Dreyer}},\ and\ \bibinfo {author} {\bibfnamefont
  {A.~N.}\ \bibnamefont {Pasupathy}},\ }\bibfield  {title} {\bibinfo {title}
  {Visualization of moir{\'e}superlattices},\ }\href
  {https://doi.org/10.1038/s41565-020-0708-3} {\bibfield  {journal} {\bibinfo
  {journal} {Nature Nanotechnology}\ }\textbf {\bibinfo {volume} {15}},\
  \bibinfo {pages} {580} (\bibinfo {year} {2020})}\BibitemShut {NoStop}%
\bibitem [{\citenamefont {Sunku}\ \emph {et~al.}(2020)\citenamefont {Sunku},
  \citenamefont {McLeod}, \citenamefont {Stauber}, \citenamefont {Yoo},
  \citenamefont {Halbertal}, \citenamefont {Ni}, \citenamefont {Sternbach},
  \citenamefont {Jiang}, \citenamefont {Taniguchi}, \citenamefont {Watanabe},
  \citenamefont {Kim}, \citenamefont {Fogler},\ and\ \citenamefont
  {Basov}}]{n19}%
  \BibitemOpen
  \bibfield  {author} {\bibinfo {author} {\bibfnamefont {S.~S.}\ \bibnamefont
  {Sunku}}, \bibinfo {author} {\bibfnamefont {A.~S.}\ \bibnamefont {McLeod}},
  \bibinfo {author} {\bibfnamefont {T.}~\bibnamefont {Stauber}}, \bibinfo
  {author} {\bibfnamefont {H.}~\bibnamefont {Yoo}}, \bibinfo {author}
  {\bibfnamefont {D.}~\bibnamefont {Halbertal}}, \bibinfo {author}
  {\bibfnamefont {G.}~\bibnamefont {Ni}}, \bibinfo {author} {\bibfnamefont
  {A.}~\bibnamefont {Sternbach}}, \bibinfo {author} {\bibfnamefont {B.-Y.}\
  \bibnamefont {Jiang}}, \bibinfo {author} {\bibfnamefont {T.}~\bibnamefont
  {Taniguchi}}, \bibinfo {author} {\bibfnamefont {K.}~\bibnamefont {Watanabe}},
  \bibinfo {author} {\bibfnamefont {P.}~\bibnamefont {Kim}}, \bibinfo {author}
  {\bibfnamefont {M.~M.}\ \bibnamefont {Fogler}},\ and\ \bibinfo {author}
  {\bibfnamefont {D.~N.}\ \bibnamefont {Basov}},\ }\bibfield  {title} {\bibinfo
  {title} {Nano-photocurrent mapping of local electronic structure in twisted
  bilayer graphene},\ }\bibfield  {booktitle} {\emph {\bibinfo {booktitle}
  {Nano Letters}},\ }\href {https://doi.org/10.1021/acs.nanolett.9b04637}
  {\bibfield  {journal} {\bibinfo  {journal} {Nano Letters}\ }\textbf {\bibinfo
  {volume} {20}},\ \bibinfo {pages} {2958} (\bibinfo {year}
  {2020})}\BibitemShut {NoStop}%
\bibitem [{\citenamefont {Zondiner}\ \emph {et~al.}(2020)\citenamefont
  {Zondiner}, \citenamefont {Rozen}, \citenamefont {Rodan-Legrain},
  \citenamefont {Cao}, \citenamefont {Queiroz}, \citenamefont {Taniguchi},
  \citenamefont {Watanabe}, \citenamefont {Oreg}, \citenamefont {von Oppen},
  \citenamefont {Stern}, \citenamefont {Berg}, \citenamefont
  {Jarillo-Herrero},\ and\ \citenamefont {Ilani}}]{n20}%
  \BibitemOpen
  \bibfield  {author} {\bibinfo {author} {\bibfnamefont {U.}~\bibnamefont
  {Zondiner}}, \bibinfo {author} {\bibfnamefont {A.}~\bibnamefont {Rozen}},
  \bibinfo {author} {\bibfnamefont {D.}~\bibnamefont {Rodan-Legrain}}, \bibinfo
  {author} {\bibfnamefont {Y.}~\bibnamefont {Cao}}, \bibinfo {author}
  {\bibfnamefont {R.}~\bibnamefont {Queiroz}}, \bibinfo {author} {\bibfnamefont
  {T.}~\bibnamefont {Taniguchi}}, \bibinfo {author} {\bibfnamefont
  {K.}~\bibnamefont {Watanabe}}, \bibinfo {author} {\bibfnamefont
  {Y.}~\bibnamefont {Oreg}}, \bibinfo {author} {\bibfnamefont {F.}~\bibnamefont
  {von Oppen}}, \bibinfo {author} {\bibfnamefont {A.}~\bibnamefont {Stern}},
  \bibinfo {author} {\bibfnamefont {E.}~\bibnamefont {Berg}}, \bibinfo {author}
  {\bibfnamefont {P.}~\bibnamefont {Jarillo-Herrero}},\ and\ \bibinfo {author}
  {\bibfnamefont {S.}~\bibnamefont {Ilani}},\ }\bibfield  {title} {\bibinfo
  {title} {Cascade of phase transitions and dirac revivals in magic-angle
  graphene},\ }\href {https://doi.org/10.1038/s41586-020-2373-y} {\bibfield
  {journal} {\bibinfo  {journal} {Nature}\ }\textbf {\bibinfo {volume} {582}},\
  \bibinfo {pages} {203} (\bibinfo {year} {2020})}\BibitemShut {NoStop}%
\bibitem [{\citenamefont {Choi}\ \emph {et~al.}(2019)\citenamefont {Choi},
  \citenamefont {Kemmer}, \citenamefont {Peng}, \citenamefont {Thomson},
  \citenamefont {Arora}, \citenamefont {Polski}, \citenamefont {Zhang},
  \citenamefont {Ren}, \citenamefont {Alicea}, \citenamefont {Refael},
  \citenamefont {von Oppen}, \citenamefont {Watanabe}, \citenamefont
  {Taniguchi},\ and\ \citenamefont {Nadj-Perge}}]{n21}%
  \BibitemOpen
  \bibfield  {author} {\bibinfo {author} {\bibfnamefont {Y.}~\bibnamefont
  {Choi}}, \bibinfo {author} {\bibfnamefont {J.}~\bibnamefont {Kemmer}},
  \bibinfo {author} {\bibfnamefont {Y.}~\bibnamefont {Peng}}, \bibinfo {author}
  {\bibfnamefont {A.}~\bibnamefont {Thomson}}, \bibinfo {author} {\bibfnamefont
  {H.}~\bibnamefont {Arora}}, \bibinfo {author} {\bibfnamefont
  {R.}~\bibnamefont {Polski}}, \bibinfo {author} {\bibfnamefont
  {Y.}~\bibnamefont {Zhang}}, \bibinfo {author} {\bibfnamefont
  {H.}~\bibnamefont {Ren}}, \bibinfo {author} {\bibfnamefont {J.}~\bibnamefont
  {Alicea}}, \bibinfo {author} {\bibfnamefont {G.}~\bibnamefont {Refael}},
  \bibinfo {author} {\bibfnamefont {F.}~\bibnamefont {von Oppen}}, \bibinfo
  {author} {\bibfnamefont {K.}~\bibnamefont {Watanabe}}, \bibinfo {author}
  {\bibfnamefont {T.}~\bibnamefont {Taniguchi}},\ and\ \bibinfo {author}
  {\bibfnamefont {S.}~\bibnamefont {Nadj-Perge}},\ }\bibfield  {title}
  {\bibinfo {title} {Electronic correlations in twisted bilayer graphene near
  the magic angle},\ }\href {https://doi.org/10.1038/s41567-019-0606-5}
  {\bibfield  {journal} {\bibinfo  {journal} {Nature Physics}\ }\textbf
  {\bibinfo {volume} {15}},\ \bibinfo {pages} {1174} (\bibinfo {year}
  {2019})}\BibitemShut {NoStop}%
\bibitem [{\citenamefont {Kerelsky}\ \emph {et~al.}(2019)\citenamefont
  {Kerelsky}, \citenamefont {McGilly}, \citenamefont {Kennes}, \citenamefont
  {Xian}, \citenamefont {Yankowitz}, \citenamefont {Chen}, \citenamefont
  {Watanabe}, \citenamefont {Taniguchi}, \citenamefont {Hone}, \citenamefont
  {Dean}, \citenamefont {Rubio},\ and\ \citenamefont {Pasupathy}}]{n22}%
  \BibitemOpen
  \bibfield  {author} {\bibinfo {author} {\bibfnamefont {A.}~\bibnamefont
  {Kerelsky}}, \bibinfo {author} {\bibfnamefont {L.~J.}\ \bibnamefont
  {McGilly}}, \bibinfo {author} {\bibfnamefont {D.~M.}\ \bibnamefont {Kennes}},
  \bibinfo {author} {\bibfnamefont {L.}~\bibnamefont {Xian}}, \bibinfo {author}
  {\bibfnamefont {M.}~\bibnamefont {Yankowitz}}, \bibinfo {author}
  {\bibfnamefont {S.}~\bibnamefont {Chen}}, \bibinfo {author} {\bibfnamefont
  {K.}~\bibnamefont {Watanabe}}, \bibinfo {author} {\bibfnamefont
  {T.}~\bibnamefont {Taniguchi}}, \bibinfo {author} {\bibfnamefont
  {J.}~\bibnamefont {Hone}}, \bibinfo {author} {\bibfnamefont {C.}~\bibnamefont
  {Dean}}, \bibinfo {author} {\bibfnamefont {A.}~\bibnamefont {Rubio}},\ and\
  \bibinfo {author} {\bibfnamefont {A.~N.}\ \bibnamefont {Pasupathy}},\
  }\bibfield  {title} {\bibinfo {title} {Maximized electron interactions at the
  magic angle in twisted bilayer graphene},\ }\href
  {https://doi.org/10.1038/s41586-019-1431-9} {\bibfield  {journal} {\bibinfo
  {journal} {Nature}\ }\textbf {\bibinfo {volume} {572}},\ \bibinfo {pages}
  {95} (\bibinfo {year} {2019})}\BibitemShut {NoStop}%
\bibitem [{\citenamefont {Xie}\ \emph {et~al.}(2019)\citenamefont {Xie},
  \citenamefont {Lian}, \citenamefont {J{\"a}ck}, \citenamefont {Liu},
  \citenamefont {Chiu}, \citenamefont {Watanabe}, \citenamefont {Taniguchi},
  \citenamefont {Bernevig},\ and\ \citenamefont {Yazdani}}]{n23}%
  \BibitemOpen
  \bibfield  {author} {\bibinfo {author} {\bibfnamefont {Y.}~\bibnamefont
  {Xie}}, \bibinfo {author} {\bibfnamefont {B.}~\bibnamefont {Lian}}, \bibinfo
  {author} {\bibfnamefont {B.}~\bibnamefont {J{\"a}ck}}, \bibinfo {author}
  {\bibfnamefont {X.}~\bibnamefont {Liu}}, \bibinfo {author} {\bibfnamefont
  {C.-L.}\ \bibnamefont {Chiu}}, \bibinfo {author} {\bibfnamefont
  {K.}~\bibnamefont {Watanabe}}, \bibinfo {author} {\bibfnamefont
  {T.}~\bibnamefont {Taniguchi}}, \bibinfo {author} {\bibfnamefont {B.~A.}\
  \bibnamefont {Bernevig}},\ and\ \bibinfo {author} {\bibfnamefont
  {A.}~\bibnamefont {Yazdani}},\ }\bibfield  {title} {\bibinfo {title}
  {Spectroscopic signatures of many-body correlations in magic-angle twisted
  bilayer graphene},\ }\href {https://doi.org/10.1038/s41586-019-1422-x}
  {\bibfield  {journal} {\bibinfo  {journal} {Nature}\ }\textbf {\bibinfo
  {volume} {572}},\ \bibinfo {pages} {101} (\bibinfo {year}
  {2019})}\BibitemShut {NoStop}%
\bibitem [{\citenamefont {Wong}\ \emph {et~al.}(2020)\citenamefont {Wong},
  \citenamefont {Nuckolls}, \citenamefont {Oh}, \citenamefont {Lian},
  \citenamefont {Xie}, \citenamefont {Jeon}, \citenamefont {Watanabe},
  \citenamefont {Taniguchi}, \citenamefont {Bernevig},\ and\ \citenamefont
  {Yazdani}}]{n24}%
  \BibitemOpen
  \bibfield  {author} {\bibinfo {author} {\bibfnamefont {D.}~\bibnamefont
  {Wong}}, \bibinfo {author} {\bibfnamefont {K.~P.}\ \bibnamefont {Nuckolls}},
  \bibinfo {author} {\bibfnamefont {M.}~\bibnamefont {Oh}}, \bibinfo {author}
  {\bibfnamefont {B.}~\bibnamefont {Lian}}, \bibinfo {author} {\bibfnamefont
  {Y.}~\bibnamefont {Xie}}, \bibinfo {author} {\bibfnamefont {S.}~\bibnamefont
  {Jeon}}, \bibinfo {author} {\bibfnamefont {K.}~\bibnamefont {Watanabe}},
  \bibinfo {author} {\bibfnamefont {T.}~\bibnamefont {Taniguchi}}, \bibinfo
  {author} {\bibfnamefont {B.~A.}\ \bibnamefont {Bernevig}},\ and\ \bibinfo
  {author} {\bibfnamefont {A.}~\bibnamefont {Yazdani}},\ }\bibfield  {title}
  {\bibinfo {title} {Cascade of electronic transitions in magic-angle twisted
  bilayer graphene},\ }\href {https://doi.org/10.1038/s41586-020-2339-0}
  {\bibfield  {journal} {\bibinfo  {journal} {Nature}\ }\textbf {\bibinfo
  {volume} {582}},\ \bibinfo {pages} {198} (\bibinfo {year}
  {2020})}\BibitemShut {NoStop}%
\bibitem [{\citenamefont {Jiang}\ \emph {et~al.}(2019)\citenamefont {Jiang},
  \citenamefont {Lai}, \citenamefont {Watanabe}, \citenamefont {Taniguchi},
  \citenamefont {Haule}, \citenamefont {Mao},\ and\ \citenamefont
  {Andrei}}]{n25}%
  \BibitemOpen
  \bibfield  {author} {\bibinfo {author} {\bibfnamefont {Y.}~\bibnamefont
  {Jiang}}, \bibinfo {author} {\bibfnamefont {X.}~\bibnamefont {Lai}}, \bibinfo
  {author} {\bibfnamefont {K.}~\bibnamefont {Watanabe}}, \bibinfo {author}
  {\bibfnamefont {T.}~\bibnamefont {Taniguchi}}, \bibinfo {author}
  {\bibfnamefont {K.}~\bibnamefont {Haule}}, \bibinfo {author} {\bibfnamefont
  {J.}~\bibnamefont {Mao}},\ and\ \bibinfo {author} {\bibfnamefont {E.~Y.}\
  \bibnamefont {Andrei}},\ }\bibfield  {title} {\bibinfo {title} {Charge order
  and broken rotational symmetry in magic-angle twisted bilayer graphene},\
  }\href {https://doi.org/10.1038/s41586-019-1460-4} {\bibfield  {journal}
  {\bibinfo  {journal} {Nature}\ }\textbf {\bibinfo {volume} {573}},\ \bibinfo
  {pages} {91} (\bibinfo {year} {2019})}\BibitemShut {NoStop}%
\bibitem [{\citenamefont {Nuckolls}\ \emph {et~al.}(2020)\citenamefont
  {Nuckolls}, \citenamefont {Oh}, \citenamefont {Wong}, \citenamefont {Lian},
  \citenamefont {Watanabe}, \citenamefont {Taniguchi}, \citenamefont
  {Bernevig},\ and\ \citenamefont {Yazdani}}]{n26}%
  \BibitemOpen
  \bibfield  {author} {\bibinfo {author} {\bibfnamefont {K.~P.}\ \bibnamefont
  {Nuckolls}}, \bibinfo {author} {\bibfnamefont {M.}~\bibnamefont {Oh}},
  \bibinfo {author} {\bibfnamefont {D.}~\bibnamefont {Wong}}, \bibinfo {author}
  {\bibfnamefont {B.}~\bibnamefont {Lian}}, \bibinfo {author} {\bibfnamefont
  {K.}~\bibnamefont {Watanabe}}, \bibinfo {author} {\bibfnamefont
  {T.}~\bibnamefont {Taniguchi}}, \bibinfo {author} {\bibfnamefont {B.~A.}\
  \bibnamefont {Bernevig}},\ and\ \bibinfo {author} {\bibfnamefont
  {A.}~\bibnamefont {Yazdani}},\ }\href@noop {} {\bibinfo {title} {Strongly
  correlated chern insulators in magic-angle twisted bilayer graphene}}
  (\bibinfo {year} {2020}),\ \Eprint {https://arxiv.org/abs/2007.03810}
  {arXiv:2007.03810 [cond-mat.mes-hall]} \BibitemShut {NoStop}%
\bibitem [{\citenamefont {Li}\ \emph {et~al.}(2011)\citenamefont {Li},
  \citenamefont {Luican},\ and\ \citenamefont {Andrei}}]{n27}%
  \BibitemOpen
  \bibfield  {author} {\bibinfo {author} {\bibfnamefont {G.}~\bibnamefont
  {Li}}, \bibinfo {author} {\bibfnamefont {A.}~\bibnamefont {Luican}},\ and\
  \bibinfo {author} {\bibfnamefont {E.~Y.}\ \bibnamefont {Andrei}},\ }\bibfield
   {title} {\bibinfo {title} {Self-navigation of a scanning tunneling
  microscope tip toward a micron-sized graphene sample},\ }\href
  {https://doi.org/10.1063/1.3605664} {\bibfield  {journal} {\bibinfo
  {journal} {Review of Scientific Instruments}\ }\textbf {\bibinfo {volume}
  {82}},\ \bibinfo {pages} {073701} (\bibinfo {year} {2011})}\BibitemShut
  {NoStop}%
\bibitem [{\citenamefont {H\"ytch}(1997)}]{n28}%
  \BibitemOpen
  \bibfield  {author} {\bibinfo {author} {\bibfnamefont {M.}~\bibnamefont
  {H\"ytch}},\ }\bibfield  {title} {\bibinfo {title} {Geometric phase analysis
  of high resolution electron microscope images},\ }\href@noop {} {\bibfield
  {journal} {\bibinfo  {journal} {Scanning Microscopy}\ }\textbf {\bibinfo
  {volume} {11}},\ \bibinfo {pages} {53} (\bibinfo {year} {1997})}\BibitemShut
  {NoStop}%
\bibitem [{\citenamefont {Zhu}\ \emph {et~al.}(2013)\citenamefont {Zhu},
  \citenamefont {Ophus}, \citenamefont {Ciston},\ and\ \citenamefont
  {Wang}}]{n29}%
  \BibitemOpen
  \bibfield  {author} {\bibinfo {author} {\bibfnamefont {Y.}~\bibnamefont
  {Zhu}}, \bibinfo {author} {\bibfnamefont {C.}~\bibnamefont {Ophus}}, \bibinfo
  {author} {\bibfnamefont {J.}~\bibnamefont {Ciston}},\ and\ \bibinfo {author}
  {\bibfnamefont {H.}~\bibnamefont {Wang}},\ }\bibfield  {title} {\bibinfo
  {title} {Interface lattice displacement measurement to 1pm by geometric phase
  analysis on aberration-corrected haadf stem images},\ }\href
  {https://doi.org/https://doi.org/10.1016/j.actamat.2013.06.006} {\bibfield
  {journal} {\bibinfo  {journal} {Acta Materialia}\ }\textbf {\bibinfo {volume}
  {61}},\ \bibinfo {pages} {5646 } (\bibinfo {year} {2013})}\BibitemShut
  {NoStop}%
\bibitem [{\citenamefont {Zhang}\ \emph {et~al.}(2016)\citenamefont {Zhang},
  \citenamefont {Liu}, \citenamefont {Wen}, \citenamefont {Xie},\ and\
  \citenamefont {Liu}}]{n30}%
  \BibitemOpen
  \bibfield  {author} {\bibinfo {author} {\bibfnamefont {H.}~\bibnamefont
  {Zhang}}, \bibinfo {author} {\bibfnamefont {Z.}~\bibnamefont {Liu}}, \bibinfo
  {author} {\bibfnamefont {H.}~\bibnamefont {Wen}}, \bibinfo {author}
  {\bibfnamefont {H.}~\bibnamefont {Xie}},\ and\ \bibinfo {author}
  {\bibfnamefont {C.}~\bibnamefont {Liu}},\ }\bibfield  {title} {\bibinfo
  {title} {Subset geometric phase analysis method for deformation evaluation of
  hrtem images},\ }\href
  {https://doi.org/https://doi.org/10.1016/j.ultramic.2016.08.019} {\bibfield
  {journal} {\bibinfo  {journal} {Ultramicroscopy}\ }\textbf {\bibinfo {volume}
  {171}},\ \bibinfo {pages} {34 } (\bibinfo {year} {2016})}\BibitemShut
  {NoStop}%
\bibitem [{\citenamefont {Kemao}(2007{\natexlab{a}})}]{n31}%
  \BibitemOpen
  \bibfield  {author} {\bibinfo {author} {\bibfnamefont {Q.}~\bibnamefont
  {Kemao}},\ }\bibfield  {title} {\bibinfo {title} {Two-dimensional windowed
  fourier transform for fringe pattern analysis: Principles, applications and
  implementations},\ }\href
  {https://doi.org/https://doi.org/10.1016/j.optlaseng.2005.10.012} {\bibfield
  {journal} {\bibinfo  {journal} {Optics and Lasers in Engineering}\ }\textbf
  {\bibinfo {volume} {45}},\ \bibinfo {pages} {304 } (\bibinfo {year}
  {2007}{\natexlab{a}})}\BibitemShut {NoStop}%
\bibitem [{\citenamefont {Lawler}\ \emph {et~al.}(2010)\citenamefont {Lawler},
  \citenamefont {Fujita}, \citenamefont {Lee}, \citenamefont {Schmidt},
  \citenamefont {Kohsaka}, \citenamefont {Kim}, \citenamefont {Eisaki},
  \citenamefont {Uchida}, \citenamefont {Davis}, \citenamefont {Sethna},\ and\
  \citenamefont {Kim}}]{n32}%
  \BibitemOpen
  \bibfield  {author} {\bibinfo {author} {\bibfnamefont {M.~J.}\ \bibnamefont
  {Lawler}}, \bibinfo {author} {\bibfnamefont {K.}~\bibnamefont {Fujita}},
  \bibinfo {author} {\bibfnamefont {J.}~\bibnamefont {Lee}}, \bibinfo {author}
  {\bibfnamefont {A.~R.}\ \bibnamefont {Schmidt}}, \bibinfo {author}
  {\bibfnamefont {Y.}~\bibnamefont {Kohsaka}}, \bibinfo {author} {\bibfnamefont
  {C.~K.}\ \bibnamefont {Kim}}, \bibinfo {author} {\bibfnamefont
  {H.}~\bibnamefont {Eisaki}}, \bibinfo {author} {\bibfnamefont
  {S.}~\bibnamefont {Uchida}}, \bibinfo {author} {\bibfnamefont {J.~C.}\
  \bibnamefont {Davis}}, \bibinfo {author} {\bibfnamefont {J.~P.}\ \bibnamefont
  {Sethna}},\ and\ \bibinfo {author} {\bibfnamefont {E.-A.}\ \bibnamefont
  {Kim}},\ }\bibfield  {title} {\bibinfo {title} {Intra-unit-cell electronic
  nematicity of the high-tc copper-oxide pseudogap states},\ }\href
  {https://doi.org/10.1038/nature09169} {\bibfield  {journal} {\bibinfo
  {journal} {Nature}\ }\textbf {\bibinfo {volume} {466}},\ \bibinfo {pages}
  {347} (\bibinfo {year} {2010})}\BibitemShut {NoStop}%
\bibitem [{\citenamefont {Slezak}\ \emph {et~al.}(2008)\citenamefont {Slezak},
  \citenamefont {Lee}, \citenamefont {Wang}, \citenamefont {McElroy},
  \citenamefont {Fujita}, \citenamefont {Andersen}, \citenamefont {Hirschfeld},
  \citenamefont {Eisaki}, \citenamefont {Uchida},\ and\ \citenamefont
  {Davis}}]{n33}%
  \BibitemOpen
  \bibfield  {author} {\bibinfo {author} {\bibfnamefont {J.~A.}\ \bibnamefont
  {Slezak}}, \bibinfo {author} {\bibfnamefont {J.}~\bibnamefont {Lee}},
  \bibinfo {author} {\bibfnamefont {M.}~\bibnamefont {Wang}}, \bibinfo {author}
  {\bibfnamefont {K.}~\bibnamefont {McElroy}}, \bibinfo {author} {\bibfnamefont
  {K.}~\bibnamefont {Fujita}}, \bibinfo {author} {\bibfnamefont {B.~M.}\
  \bibnamefont {Andersen}}, \bibinfo {author} {\bibfnamefont {P.~J.}\
  \bibnamefont {Hirschfeld}}, \bibinfo {author} {\bibfnamefont
  {H.}~\bibnamefont {Eisaki}}, \bibinfo {author} {\bibfnamefont
  {S.}~\bibnamefont {Uchida}},\ and\ \bibinfo {author} {\bibfnamefont {J.~C.}\
  \bibnamefont {Davis}},\ }\bibfield  {title} {\bibinfo {title} {Imaging the
  impact on cuprate superconductivity of varying the interatomic distances
  within individual crystal unit cells},\ }\href
  {https://doi.org/10.1073/pnas.0706795105} {\bibfield  {journal} {\bibinfo
  {journal} {Proceedings of the National Academy of Sciences}\ }\textbf
  {\bibinfo {volume} {105}},\ \bibinfo {pages} {3203} (\bibinfo {year}
  {2008})}\BibitemShut {NoStop}%
\bibitem [{\citenamefont {Bi}\ \emph {et~al.}(2019)\citenamefont {Bi},
  \citenamefont {Yuan},\ and\ \citenamefont {Fu}}]{n34}%
  \BibitemOpen
  \bibfield  {author} {\bibinfo {author} {\bibfnamefont {Z.}~\bibnamefont
  {Bi}}, \bibinfo {author} {\bibfnamefont {N.~F.~Q.}\ \bibnamefont {Yuan}},\
  and\ \bibinfo {author} {\bibfnamefont {L.}~\bibnamefont {Fu}},\ }\bibfield
  {title} {\bibinfo {title} {Designing flat bands by strain},\ }\href
  {https://doi.org/10.1103/PhysRevB.100.035448} {\bibfield  {journal} {\bibinfo
   {journal} {Phys. Rev. B}\ }\textbf {\bibinfo {volume} {100}},\ \bibinfo
  {pages} {035448} (\bibinfo {year} {2019})}\BibitemShut {NoStop}%
\bibitem [{\citenamefont {Li}\ \emph {et~al.}(2010)\citenamefont {Li},
  \citenamefont {Luican}, \citenamefont {Lopes~dos Santos}, \citenamefont
  {Castro~Neto}, \citenamefont {Reina}, \citenamefont {Kong},\ and\
  \citenamefont {Andrei}}]{n35}%
  \BibitemOpen
  \bibfield  {author} {\bibinfo {author} {\bibfnamefont {G.}~\bibnamefont
  {Li}}, \bibinfo {author} {\bibfnamefont {A.}~\bibnamefont {Luican}}, \bibinfo
  {author} {\bibfnamefont {J.~M.~B.}\ \bibnamefont {Lopes~dos Santos}},
  \bibinfo {author} {\bibfnamefont {A.~H.}\ \bibnamefont {Castro~Neto}},
  \bibinfo {author} {\bibfnamefont {A.}~\bibnamefont {Reina}}, \bibinfo
  {author} {\bibfnamefont {J.}~\bibnamefont {Kong}},\ and\ \bibinfo {author}
  {\bibfnamefont {E.~Y.}\ \bibnamefont {Andrei}},\ }\bibfield  {title}
  {\bibinfo {title} {Observation of van hove singularities in twisted graphene
  layers},\ }\href {https://doi.org/10.1038/nphys1463} {\bibfield  {journal}
  {\bibinfo  {journal} {Nature Physics}\ }\textbf {\bibinfo {volume} {6}},\
  \bibinfo {pages} {109} (\bibinfo {year} {2010})}\BibitemShut {NoStop}%
\bibitem [{\citenamefont {Dutreix}\ \emph {et~al.}(2019)\citenamefont
  {Dutreix}, \citenamefont {Gonz{\'a}lez-Herrero}, \citenamefont {Brihuega},
  \citenamefont {Katsnelson}, \citenamefont {Chapelier},\ and\ \citenamefont
  {Renard}}]{n38}%
  \BibitemOpen
  \bibfield  {author} {\bibinfo {author} {\bibfnamefont {C.}~\bibnamefont
  {Dutreix}}, \bibinfo {author} {\bibfnamefont {H.}~\bibnamefont
  {Gonz{\'a}lez-Herrero}}, \bibinfo {author} {\bibfnamefont {I.}~\bibnamefont
  {Brihuega}}, \bibinfo {author} {\bibfnamefont {M.~I.}\ \bibnamefont
  {Katsnelson}}, \bibinfo {author} {\bibfnamefont {C.}~\bibnamefont
  {Chapelier}},\ and\ \bibinfo {author} {\bibfnamefont {V.~T.}\ \bibnamefont
  {Renard}},\ }\bibfield  {title} {\bibinfo {title} {Measuring the berry phase
  of graphene from wavefront dislocations in friedel oscillations},\ }\href
  {https://doi.org/10.1038/s41586-019-1613-5} {\bibfield  {journal} {\bibinfo
  {journal} {Nature}\ }\textbf {\bibinfo {volume} {574}},\ \bibinfo {pages}
  {219} (\bibinfo {year} {2019})}\BibitemShut {NoStop}%
\bibitem [{\citenamefont {Nam}\ and\ \citenamefont {Koshino}(2017)}]{n36}%
  \BibitemOpen
  \bibfield  {author} {\bibinfo {author} {\bibfnamefont {N.~N.~T.}\
  \bibnamefont {Nam}}\ and\ \bibinfo {author} {\bibfnamefont {M.}~\bibnamefont
  {Koshino}},\ }\bibfield  {title} {\bibinfo {title} {Lattice relaxation and
  energy band modulation in twisted bilayer graphene},\ }\href
  {https://doi.org/10.1103/PhysRevB.96.075311} {\bibfield  {journal} {\bibinfo
  {journal} {Phys. Rev. B}\ }\textbf {\bibinfo {volume} {96}},\ \bibinfo
  {pages} {075311} (\bibinfo {year} {2017})}\BibitemShut {NoStop}%
\bibitem [{\citenamefont {Balents}(2019)}]{n37}%
  \BibitemOpen
  \bibfield  {author} {\bibinfo {author} {\bibfnamefont {L.}~\bibnamefont
  {Balents}},\ }\bibfield  {title} {\bibinfo {title} {{General continuum model
  for twisted bilayer graphene and arbitrary smooth deformations}},\ }\href
  {https://doi.org/10.21468/SciPostPhys.7.4.048} {\bibfield  {journal}
  {\bibinfo  {journal} {SciPost Phys.}\ }\textbf {\bibinfo {volume} {7}},\
  \bibinfo {pages} {48} (\bibinfo {year} {2019})}\BibitemShut {NoStop}%
\bibitem [{\citenamefont {Ghiglia}\ and\ \citenamefont
  {Romero}(1994)}]{Ghiglia1994}%
  \BibitemOpen
  \bibfield  {author} {\bibinfo {author} {\bibfnamefont {D.~C.}\ \bibnamefont
  {Ghiglia}}\ and\ \bibinfo {author} {\bibfnamefont {L.~A.}\ \bibnamefont
  {Romero}},\ }\bibfield  {title} {\bibinfo {title} {Robust two-dimensional
  weighted and unweighted phase unwrapping that uses fast transforms and
  iterative methods},\ }\href {https://doi.org/10.1364/josaa.11.000107}
  {\bibfield  {journal} {\bibinfo  {journal} {Journal of the Optical Society of
  America A}\ }\textbf {\bibinfo {volume} {11}},\ \bibinfo {pages} {107}
  (\bibinfo {year} {1994})}\BibitemShut {NoStop}%
\bibitem [{\citenamefont {Herr{\'{a}}ez}\ \emph {et~al.}(2002)\citenamefont
  {Herr{\'{a}}ez}, \citenamefont {Burton}, \citenamefont {Lalor},\ and\
  \citenamefont {Gdeisat}}]{Herrez2002}%
  \BibitemOpen
  \bibfield  {author} {\bibinfo {author} {\bibfnamefont {M.~A.}\ \bibnamefont
  {Herr{\'{a}}ez}}, \bibinfo {author} {\bibfnamefont {D.~R.}\ \bibnamefont
  {Burton}}, \bibinfo {author} {\bibfnamefont {M.~J.}\ \bibnamefont {Lalor}},\
  and\ \bibinfo {author} {\bibfnamefont {M.~A.}\ \bibnamefont {Gdeisat}},\
  }\bibfield  {title} {\bibinfo {title} {Fast two-dimensional phase-unwrapping
  algorithm based on sorting by reliability following a noncontinuous path},\
  }\href {https://doi.org/10.1364/ao.41.007437} {\bibfield  {journal} {\bibinfo
   {journal} {Applied Optics}\ }\textbf {\bibinfo {volume} {41}},\ \bibinfo
  {pages} {7437} (\bibinfo {year} {2002})}\BibitemShut {NoStop}%
\bibitem [{\citenamefont {Kemao}(2007{\natexlab{b}})}]{Kemao2007}%
  \BibitemOpen
  \bibfield  {author} {\bibinfo {author} {\bibfnamefont {Q.}~\bibnamefont
  {Kemao}},\ }\bibfield  {title} {\bibinfo {title} {Two-dimensional windowed
  fourier transform for fringe pattern analysis: Principles, applications and
  implementations},\ }\href {https://doi.org/10.1016/j.optlaseng.2005.10.012}
  {\bibfield  {journal} {\bibinfo  {journal} {Optics and Lasers in
  Engineering}\ }\textbf {\bibinfo {volume} {45}},\ \bibinfo {pages} {304}
  (\bibinfo {year} {2007}{\natexlab{b}})}\BibitemShut {NoStop}%
\bibitem [{\citenamefont {Kasim}(2016)}]{Firman2016}%
  \BibitemOpen
  \bibfield  {author} {\bibinfo {author} {\bibfnamefont {M.~F.}\ \bibnamefont
  {Kasim}},\ }\href
  {https://www.mathworks.com/matlabcentral/fileexchange/60345-2d-weighted-phase-unwrapping}
  {\bibinfo {title} {2d weighthed phase unwrapping}} (\bibinfo {year} {2016}),\
  \bibinfo {note} {matlab central file exchange}\BibitemShut {NoStop}%
\bibitem [{\citenamefont {Moisan}(2010)}]{Moisan2010}%
  \BibitemOpen
  \bibfield  {author} {\bibinfo {author} {\bibfnamefont {L.}~\bibnamefont
  {Moisan}},\ }\bibfield  {title} {\bibinfo {title} {Periodic plus smooth image
  decomposition},\ }\href {https://doi.org/10.1007/s10851-010-0227-1}
  {\bibfield  {journal} {\bibinfo  {journal} {Journal of Mathematical Imaging
  and Vision}\ }\textbf {\bibinfo {volume} {39}},\ \bibinfo {pages} {161}
  (\bibinfo {year} {2010})}\BibitemShut {NoStop}%
\end{thebibliography}
\end{document}